\documentclass[fleqn,usenatbib]{mnras}
\usepackage{comment}

\usepackage{newtxtext,newtxmath}

\usepackage[T1]{fontenc}
\usepackage{ae,aecompl}

\usepackage{graphicx}	
\usepackage{amsmath}	
\usepackage[nameinlink,noabbrev]{cleveref}

\usepackage{xcolor}

\usepackage[separate-uncertainty=true,multi-part-units=single]{siunitx}        

\def\be{\begin{equation}}
\def\ee{\end{equation}}

\newcommand\quotes[1]{``{#1}"}
\def\gsim{\lower.5ex\hbox{\gtsima}} 
\def\lsim{\lower.5ex\hbox{\ltsima}} 
\def\gtsima{$\; \buildrel > \over \sim \;$} 
\def\ltsima{$\; \buildrel < \over \sim \;$} \def\gsim{\lower.5ex\hbox{\gtsima}} 
\def\lsim{\lower.5ex\hbox{\ltsima}} 
\def\simgt{\lower.5ex\hbox{\gtsima}} 
\def\simlt{\lower.5ex\hbox{\ltsima}}

\def\msun{{\rm M}_{\odot}}

\newcommand{\angstrom}{\mbox{\normalfont\AA}}
\def\cc{\rm cm^{-3}}

\def\S*{$\Sigma_{\rm SFR}$}

\def\Lya{Ly$\alpha$}
 
\def\HI{\hbox{H~$\scriptstyle\rm I\ $}} 
\def\HII{\hbox{H~$\scriptstyle\rm II\ $}}

\definecolor{apcolor}{HTML}{b3003b}
\definecolor{afcolor}{HTML}{800080}
\definecolor{lvcolor}{HTML}{DF7401}
\definecolor{mdcolor}{HTML}{01abdf} 
\definecolor{cbcolor}{HTML}{ff0000}
\definecolor{sccolor}{HTML}{cc5500} 
\definecolor{sgcolor}{HTML}{00cc7a}

\newcommand{\dd}{\mathop{\mathrm{d}\!}{}}
\newcommand{\ped}[1]{_{\mathrm{#1}}}
\newcommand{\ap}[1]{^{\mathrm{#1}}}
\renewcommand{\vec}[1]{\boldsymbol{\mathbf{#1}}}
\newcommand{\ver}[1]{\vec{\hat #1}}
\DeclareMathOperator{\artanh}{artanh}
\DeclareMathOperator{\Li}{Li}

\title[Ly$\alpha$ radiation pressure]{Lyman-alpha radiation pressure: an analytical exploration}

\author[Tomaselli \& Ferrara]{G. M. Tomaselli$^{1,2}$\thanks{E-mail: giovanni.tomaselli@sns.it}, A. Ferrara$^{1}$\\
$^{1}$Scuola Normale Superiore, Piazza dei Cavalieri 7, 56126 Pisa, Italy\\
$^{2}$GRAPPA,
Institute for Theoretical Physics, University of Amsterdam, Science Park 904, 1098 XH Amsterdam, The Netherlands\\
}

\date{Accepted XXX. Received YYY; in original form ZZZ}

\pubyear{2020}

\begin{document}
\label{firstpage}
\pagerange{\pageref{firstpage}--\pageref{lastpage}}
\maketitle

\begin{abstract}
We study radiation pressure due to \Lya\ line photons, obtaining and exploring analytical expressions for the force-multiplier, $M\ped{F}(N\ped{H}, Z)=F_\alpha/(L_\alpha/c)$, as a function of gas column density, $N\ped{H}$, and metallicity, $Z$, for both dust-free and dusty media, employing a WKB approach for the latter case. Solutions for frequency offset emission to emulate non-static media moving with a bulk velocity $v$, have also been obtained. We find that, in static media, \Lya\ pressure dominates over both photoionization and dust-mediated UV radiation pressure in a very wide parameter range ($16 < \log N\ped{H} < 23; -4 < \log [Z/Z_\odot] < 0$). For example, it overwhelms the other two forces by $\simgt 10$ (300) times in standard (low-$Z$) star-forming clouds. Thus, in agreement with previous studies, we conclude that \Lya\ pressure plays a dominant role in the initial acceleration of the gas around luminous sources, and must be implemented in galaxy formation, evolution and outflow models and simulations.    
\end{abstract}

\section{Introduction}
The hydrogen Lyman Alpha (\Lya) line is often the most prominent emission feature \citep{Osterbrock06} seen in the spectra of galaxies. For this reason it has been historically \citep{Partridge67, Djorgovski92,Rhoads00,Taniguchi05} used as a primary tool to investigate the physics of star-forming systems up to high redshifts \citep[][]{Ouchi09,Hu10,Pentericci11,Kashikawa11,Ouchi18,Shibuya18}. For a complete review we refer the interested reader to \citet{Dayal18,bookDijkstra19,Ouchi:2020zce}. 

Recent dedicated surveys, such as the Hobby-Eberly Telescope Dark Energy Experiment (HETDEX) \citep{Hetdex16}, the Multi Unit Spectroscopic Explorer (VLT/MUSE) Wide Survey \citep{MUSE19}, and the Systematic Identification of LAEs for Visible
Exploration and Reionization Research Using
Subaru HSC (SILVERRUSH) \citep{Ouchi18, Shibuya18, Shibuya2019},  using narrow-band filters specifically tuned to match the redshifted \Lya\, ($\lambda_\alpha=\SI{ 1215.67}{\angstrom}$) wavelength or Integral Field Spectroscopy, have discovered a vast population of so-called Lyman Alpha Emitters (LAEs). The SILVERRUSH, in particular, has pushed the search well into the Epoch of Reionization extending through the first cosmic billion years (redshift $z>6$). The line detection has provided crucial and unique information on the ionization state of the intergalactic medium (IGM) which determines the line transmissivity. In turn, these data have allowed to trace the reionization progress with unprecedented precision \citep[e.g.][]{Dijkstra14, Mesinger15, Sobacchi15, Weinberger19, Whitler20, Jung20}.        

In spite of these manifest advantages, inferring the physical properties of galaxies from the information encoded in the line has proven extremely challenging. Although the emission physics is very well understood from basic quantum mechanical principles, the bottleneck resides in the complex description of the radiative transfer effects \citep[e.g.][]{Smith18,Behrens19,Laursen19,Hayes20,Li20}. \Lya\, photons undergo resonant scattering with neutral hydrogen (\HI) atoms performing a random walk in both frequency and space as they travel in the interstellar (ISM) or circumgalactic medium (CGM). In addition, if dust grains are present, which is almost always the case in observationally accessible environments, they can absorb or scatter \Lya\, photons. These processes depend on the gas column density, dynamics, geometry, inhomogeneities and dust content. As such, they dramatically affect the emerging line intensity and profile, making the reverse-engineering procedures, necessary to trace back the physical parameters from observations, quite arduous. 

In spite of the monumental amount of work devoted to \Lya\ visibility in the literature, comparatively much less attention has been dedicated to its dynamical effects. As the line carries a sizeable fraction of the bolometric luminosity of the galaxy, one expects that it can induce important radiation pressure effects on the surrounding gas, particularly in low-metallicity environments, where dust and heavy elements provide only a limited opacity to UV photons. 

A paradigmatic example concerns the first metal-free (PopIII) stars. Using the analytical framework laid down by the pivotal papers \citet{Adams72,Harrington73,Neufeld90}, \citet{Tan03} and \citet{McKee:2007yx}
studied the impact of \Lya\ on the infall onto the protostar, concluding that the mechanism can reduce the efficiency of the accretion process, although not completely stop it. Similarly, \citet{Oh:2001ex} analyzed the effect of \Lya\ feedback in haloes with virial temperature above $\SI{e+4}{\kelvin}$, finding a significant slow-down of the collapse.

The idea that Ly$\alpha$ pressure could represent a significant radiative feedback source for early structures was further explored by several authors: \citet{Cox85,Bithell90} and, in particular, \citet{Dijkstra08}. The last authors, with the help of a Monte Carlo radiative transfer code, showed that the force exerted by \Lya\ photons on H-atoms in early galaxies may exceed gravity by orders of magnitude, and drive supersonic winds that reach hundreds of kilometres per second. In a follow-up study, the same authors showed that \Lya\ radiation pressure provides a suitable explanation for the outflowing supershells observed around star-forming galaxies \citet{Dijkstra09}. They quantify the boost deriving from multiple scatterings of \Lya\ photons in the gas defining a ``force multiplier'', $M_{\rm F}$, one of the central aspects of this work. Physically, $M_{\rm F}$ represents the ratio of the trapping time (due to multiple scatterings) of \Lya\ photons to the light crossing time in a system of characteristic size $R$: $M_{\rm F}=t_{\rm trap}/(R/c)$.

The physics of \Lya\ transfer has been discussed in a series of papers by \citet{Smith:2006,Smith17,Smith:2017xwu, Smith18, Smith19}. In the first two they perform 1D radiation-hydrodynamics simulations of \Lya\ radiation-driven shells in early, metal-free minihalos (dark matter halos with virial temperatures $< 10^4$ K presumably hosting the first stars) and atomic cooling haloes. Adopting a Monte Carlo radiative transfer code (MCRT) coupled to a hydrodynamic solver, they conclude that \Lya\ radiation pressure can be dynamically important for a number of realistic protogalaxy environments, and accelerates the gas to velocities $2-3$ times larger than obtained from the momentum injection of ionizing radiation alone. In the subsequent works, these results are implemented in cosmological zoom-in galaxy simulations to assess the impact of \Lya\ radiation pressure on cosmic structures and the formation of direct collapse black holes. All these results agree that \Lya\ pressure is a strictly necessary ingredient of reliable models.

A proper, on-the-fly treatment of \Lya\ radiative transfer in cosmological simulations poses tremendous computational difficulties. For example, although \citet{Hopkins20} performed a very detailed study of a number of radiative feedback processes, including photoionization, photoelectric, and Compton heating, and radiation pressure on dust, they could not include the \Lya\ pressure contribution. This is because \Lya\ scattering requires custom algorithms and is extremely
computationally demanding. Hence, having a simple but exact analytical treatment which could be implemented in cosmological codes would be highly valuable.  

Recently, \citet{Kimm18} incorporated a local subgrid model for \Lya\ momentum transfer into 3D RHD simulations of an isolated metal-poor dwarf galaxy. Interestingly, the authors make the important point that \Lya\ feedback can govern the dynamics of star-forming clouds before the onset of supernova explosions, thereby suppressing or at least regulating star formation. The authors  calibrated their subgrid model by calculating $M\ped{F}$ from a Monte Carlo simulation of uniform  static spheres, based on the \textsc{RASCAS} code \citep{Michel20}, which includes recoil effects, dipolar angular redistribution functions, scattering by deuterium, and dust scattering/absorption effects. 

The fit to the numerical results shows that in the dust-free case, $M_{\rm F} \propto \tau_0^{0.29}$, where $\tau_0 =\sigma_0 N_{\rm HI}$ is the optical depth at the line center, $\sigma_0=5.88\times 10^{-14} T_4^{-1/2} \rm cm^{2}$, $T$ is the gas temperature\footnote{We use the notation $Y_X = Y/10^X.$}, and $N_{\rm HI}$ is the neutral hydrogen column density of the system. The result was not a surprise, as a power-law scaling $M\ped{F}\propto\tau_0^{1/3}$ had been predicted by \citet{Adams:1975,Smith17}. The main effect of dust is to limit the number of scatterings suffered by \Lya\ photons before they get absorbed. For this reason, at large $\tau_0$ values, $M\ped{F}$ saturates at a value which is a decreasing function of metallicity, and that for $Z= Z_\odot$ is $M\ped{F}\simeq 50$.

In this paper we aim at developing a treatment yielding a novel analytical expression for the force-multiplier depending on the various properties of the target system for which \Lya\ radiation pressure needs to be evaluated. In addition to provide a benchmark test for numerical radiative transfer simulations, our results are suitable for computationally affordable implementations in numerical simulations with applications ranging from star-forming clouds to stars, and from galaxies up to the largest cosmic structures.

\section{Optically thick line radiative transfer}
We use the general theory of radiative transfer around a resonance \citep[see, e.g.][]{Dijkstra17} to derive an analytical expression for the line radiation pressure, in the case of an optically thick medium. This problem has been tackled by a few previous works \citep{1979ApJ...233..649B,Smith17,Kimm18}, but always with the use of numerical simulations. Here instead, we intend to develop a complementary approach that is based on an analytical solution of the problem. 

The temperature of the medium plays an important role in resonant scattering, in that it broadens the line. The resulting ``thermal'' cross section is described by the Voigt profile:
\be \sigma\ped{H}=\sigma_0 H(x),\qquad x=(\nu-\nu_0)/\Delta\nu_0,\ee 
where $\sigma_0$ is the cross-section at line centre, $v_0$ and $\Delta\nu_0=\nu_0 (v\ped{th}/c)$ are the frequency and the thermal width of the resonance (in our case, the \Lya), respectively, and 
\be  H(x)=\frac{a_v}\pi\int_{-\infty}^{+\infty}\frac{e^{-y^2}\dd y}{(y-x)^2+a_v^2},\qquad a_v=4.7\times10^{-4}T_4^{-1/2}.\ee 
The shape of the Voigt profile is dominated by the thermal or natural width depending on the distance from the line center:
\be  H(x)=\begin{cases}
e^{-x^2} & \text{``core'' (small }|x|\text{),}\\
\frac{a_v}{\sqrt\pi x^2} & \text{``wing'' (large }|x|\text{).}\\
\end{cases}\ee 

The physics of time-independent radiative transfer is described by the well-known equation
\begin{equation}
\label{eqn:radiative-transfer}
\vec{\ver n\cdot\nabla}I(\vec r, \ver n, \nu)=-\alpha\ped{abs}I(\mathbf{r},\mathbf{\hat n},\nu)+j\ped{\rm source}+j\ped{\rm scatt}.
\end{equation}
Here, $I(\mathbf{r},\mathbf{\hat n},\nu)$ is the specific intensity as function of the position, direction and frequency;
the absorption coefficient is given by
$\alpha\ped{abs}=n\ped{H}\sigma_0 H(x)+n\ped{H}\sigma\ped{d}$, where $\sigma_d$ is the dust extinction cross-section in the vicinity of the \Lya\ frequency; the emissivity has been explicitly written as a sum of the contribution from external sources, $j\ped{\rm source}$, and scattering on hydrogen atoms, $j\ped{\rm scatt}$.

When the medium is extremely optically thick ($\tau_0 \gg 10^3/a_v$, see \citet{Ahn:2001pz}), photons scatter so many times before exiting the cloud that the specific intensity becomes nearly isotropic. This effect can be incorporated into \cref{eqn:radiative-transfer} using the Eddington approximation, where only the lowest-order moments of the angular dependence of $I$ are kept.

Thermal agitation causes photons to diffuse in frequency space as well, with a tendency for restoration back to the core. In the limit of large optical depth, the scattering term in \cref{eqn:radiative-transfer} can be approximated as 
\be J\ped{scatt}=n\ped{H}\sigma_0 H(x)J+n\ped{H}\sigma_0\frac{1}{2}\frac{\partial}{\partial x}\biggl( H(x)\frac{\partial J}{\partial x}\biggr),\ee 
where $J$ denotes the angle-averaged specific intensity. With these assumptions, \cref{eqn:radiative-transfer} becomes a Fokker-Planck equation. This analytical approach was developed in various works (\citet{unno1952note,Osterbrock:1962,Hummer:1962}), with \citet{Harrington73} ultimately showing for the first time explicit solutions for the specific intensity for a gas slab; more recently, \citet{Dijkstra06} generalized the solution to spherical geometry and \citet{Lao:2020ptq} extended to arbitrary density profiles. In Appendix \ref{app:computations} we review in detail the computation, with the specific goal of applying this analytical approach to the issue of the radiation pressure, and to compare our results with numerical simulations in the literature.

After some manipulations, outlined  in Appendix \ref{app:computations}, the radiative transfer equation in the case of a spherical, uniform cloud becomes
\begin{equation}
\frac{\partial^2J}{\partial \tau^2}+\frac{2}{\tau}\frac{\partial J}{\partial \tau}+\frac{\partial^2J}{\partial\sigma^2}=3 H\epsilon J-\frac{3 H}{n\ped{H}\sigma_0}J\ped{source},
\label{eqn:radtrasftausigma}
\end{equation}
where $\tau$ is the optical depth from the geometric centre and $\sigma$,
\be \sigma(x)=\int_0^x\frac{1}{\sqrt{3/2} H(y)}\dd y\xrightarrow{\text{wing (large }|x|\text{)}}\sqrt{\frac{2\pi}3}\frac{x^3}{3a_v},\ee 
is a variable in the frequency space used to conveniently rewrite the diffusion term. In addition, we have defined $\epsilon = \sigma_d/\sigma_0$. 

Once the source term and the boundary conditions are specified, \cref{eqn:radtrasftausigma} can be solved by separation of variables. This procedure deserves particular attention. While it is clear that in frequency space $J \to 0$ far away from the resonance, the treatment of the cloud edge, where directionality is restored, is far from trivial. The assumption commonly made in the literature is to enforce the Eddington approximation at the boundary; we will refer to this as the ``standard'' boundary condition.  We will compare it to an ``alternative'', flux-free, one (see Appendix \ref{app:computations} for details), to check that results on radiation pressure do not depend critically on the conditions at the edge, which only influence the physics within a ``skin depth'' of the surface. We will show that this is indeed the case, making our analytical computation robust.

\section{Analytical results}

The net radiation force on a spherical cloud is zero by spherical symmetry. However, radiation pressure will make it expand isotropically. A possible way to quantify the effect is to define
\begin{align}
F\ped{rad}&=\int_0^R4\pi r^2n\ped{H}F\ped{atom}(r)\dd r\\
&=-\frac{4\pi\Delta\nu_0}{\sqrt{6}n\ped{H}^2\sigma_0^2c}\sqrt{\frac23}\int_{-\infty}^{+\infty}\biggl(\Bigl[\tau^2J\Bigr]_0^{\tau_0}-2\int_0^{\tau_0}\tau J\dd\tau\biggr)\dd x.
\label{eqn:Frad}
\end{align}
where $R$ is the radius of the cloud, $\tau_0=n\ped{H}\sigma_0R$ its centre-to-edge optical depth and\footnote{This definition ignores recoil effects, which introduce corrections depending on the finite mass of the atom.}
\be \begin{split}
F\ped{atom}(r)&=\frac{1}{c}\int_0^\infty\dd\nu\sigma_0 H(\nu)\frac1{4\pi}\int\dd\Omega\,\vec{\ver n\cdot\ver r}\,J(\vec r,\ver n,\nu)\dd\nu\\
&\approx-\frac{\sigma_0\Delta\nu_0}{\sqrt{6}c}\int_{-\infty}^{+\infty}\frac{\partial J}{\partial\tau} H\dd\sigma,
\end{split}\ee 
is the radial force acting on a single H atom, while, for completeness, the pressure at any point in the cloud is given by
\be p\ped{rad}(r)=p\ped{rad}(0)-\int_0^r\dd r'n\ped{H}F\ped{atom}(r').\ee 

\subsection{Central source, no dust}

The first and simplest example of source is a central point source, embedded in a dust-free (corresponding to setting $\epsilon=0$ in \cref{eqn:radtrasftausigma}) medium, and emitting \Lya\ radiation isotropically. A measure of the the effect of resonant scattering on radiation pressure is given by the \quotes{force multiplier}, 
\be M\ped{F}=\frac{F\ped{rad}}{L_\alpha/c},\ee 
where $L_\alpha$ is the \Lya\ luminosity emitted by the source. As already mentioned, $M\ped{F}$ quantifies how many times, on average, a photon contributes to the momentum deposition, with respect to a  single-scattering scenario.

With standard boundary conditions, our analytical result is\footnote{We use the definition  $\artanh(x)=\frac12\log\bigl(\frac{1+x}{1-x}\bigr)$.}
\begin{equation}
\label{eqn:MFstd-central-nodust}
{M\ped{F}\ap{std}=\frac4\pi\sqrt{\frac23}\int_{-\infty}^{+\infty}\artanh\bigl(e^{-\pi|\sigma(x)|/\tau_0}\bigr)\dd x.}
\end{equation}
When $\tau_0$ is very large, photons diffuse far into the wing, therefore we may approximate $\sigma(x)$ with its large-$x$ limit to get
\begin{equation}
\label{eqn:MFstd-central-nodust-asymptotic}
{M\ped{F}\ap{std}\approx3.51\,(\tau_0a_v)^{1/3}.}
\end{equation}
This result has already been derived, independently, by \citet{Lao:2020ptq} (see their equation (92).

\begin{figure}
\includegraphics[width=0.48\textwidth]{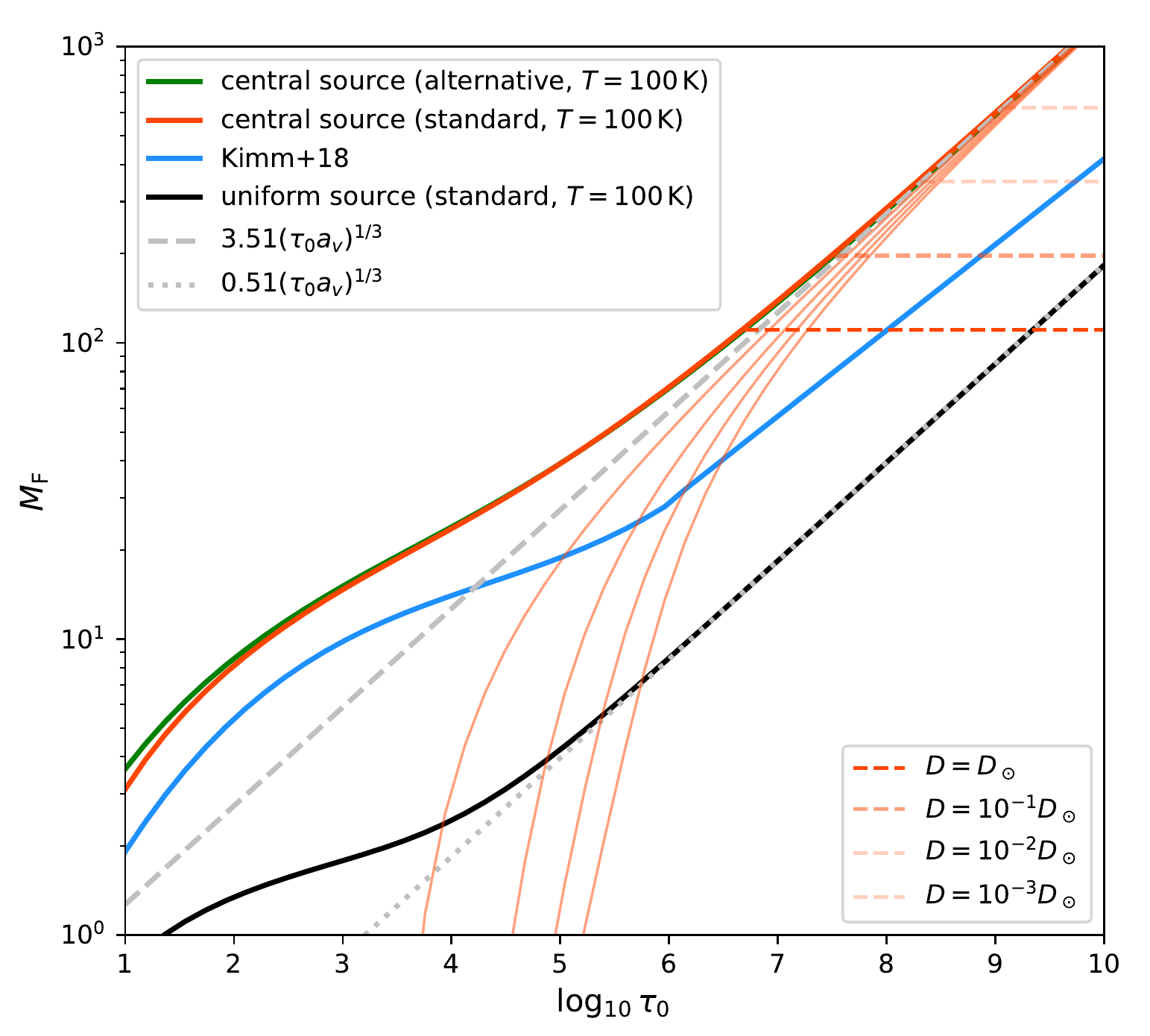}
\caption{Comparison of various estimates of the multiplication factor in the dust-free case. The thick orange, green and black solid lines are numerical integrations of \cref{eqn:MFstd-central-nodust,eqn:MF-central-alternative,eqn:MF-uniform} respectively. The dashed and dotted gray lines are the asymptotic limits \cref{eqn:MFstd-central-nodust-asymptotic} and \cref{eqn:MF-uniform-asymptotic}. The plot shows the insensitivity of $M\ped{F}$ to the choice of boundary conditions, as well as the earlier onset of the asymptotic regime in the case of uniform source. The {horizontal} orange dashed lines represent the asymptotic values of $M\ped{F}$ when dust is present, for various values of $D$, according to \cref{eqn:MFstd-central-dust}. The blue line is taken from \citet{Kimm18} and consists of two separate numerical fits to Monte-Carlo generated data at $T=\SI{100}{\kelvin}$. Although the analytical and numerical trends are the same, the two differ by a multiplicative factor $1.5-3$ in the considered $\tau_0$ interval. We have been unable to identify the origin of such discrepancy. The thin orange solid lines correspond to non-static solutions for a medium in bulk motion with velocity $v = x_{\rm s} v_{\rm th}$, with $x\ped{s}=4,8,12,16$ (see  \cref{eqn:M_F-shifted}). These solutions are discussed in Sec. \ref{sec:summary}.
\label{fig:central-nodust-comparison}} 
\end{figure}

With the alternative boundary conditions, the asymptotic limit only changes to $M\ped{F}\ap{alt}\approx3.43(\tau_0a_v)^{1/3}$. The results are plotted in Fig.~\ref{fig:central-nodust-comparison}, as well as compared with numerical estimates found in the literature.

The predictions of \citet{Kimm18} are qualitatively confirmed, including the power-law scaling at large $\tau_0$ ($1/3$ from analytical vs $0.29$ from numerical approach). The slope of 1/3 was also predicted by \citet{Adams:1975} and further explored by \citet{Smith17}, which pointed out the flatter behaviour of numerics, suggesting the inclusion of the recoil as a possible cause. Apart from the asymptotic slope, the general shape of \cref{eqn:MFstd-central-nodust} seems to be very close to \citet{Kimm18} (see Fig.~\ref{fig:central-nodust-comparison}), except for a multiplicative offset.

The effects of an expansion of the medium, with bulk velocity $v = x_{\rm s} v_{\rm th}$, can be approximately captured by inserting photons with an appropriate frequency offset\footnote{This approximation will not take into account the inevitable velocity gradients, but will capture the general behaviour.}. In this case, eq.~\ref{eqn:MFstd-central-nodust} is modified by making the transformation $\sigma(x)\to\sigma(x)-\sigma(x\ped{s})$, see eq.~\ref{eqn:M_F-shifted}. These are also shown in Fig.~\ref{fig:central-nodust-comparison}, and further discussed in Sec. \ref{sec:summary}.
%
%
\begin{figure*}
    \centering
    \includegraphics[width=0.49\linewidth]{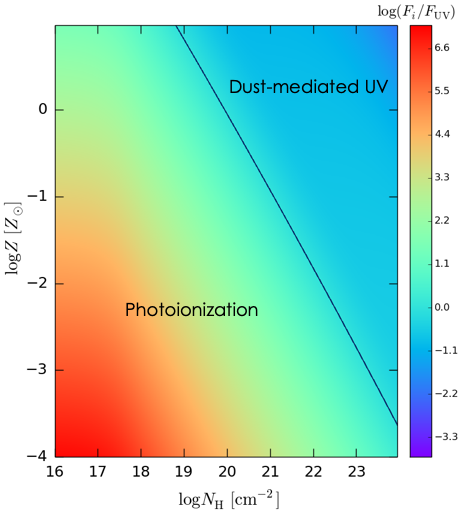}
    \includegraphics[width=0.49\linewidth]{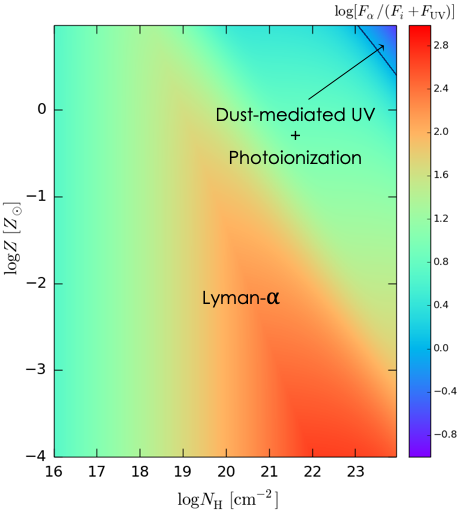}
\caption{Regimes in the gas column density, $N\ped{H}$, and metallicity, $Z$, in which different radiation forces dominate. {\it Left}: Contours show the ratio between photoionization (see Sec. \ref{sec:photo}) and dust-mediated UV (Sec. \ref{sec:UV} radiation pressure, $F_i/F_{\rm UV}$. Photoionization dominates at low $N\ped{H}$ and $Z$. The black line show the locus where the ratio is equal to 1. {\it Right}: Same for the ratio of the \Lya (Sec. \ref{sec:Lya}) to (photoionization + dust mediated UV) radiation pressure, $F_\alpha/(F_i+F_{\rm UV})$. \Lya\ pressure dominates in the entire parameter space, with exception of the tiny upper-right corner, as indicated. The black line show the locus where the ratio is equal to 1.} 
\label{Fig02}
\end{figure*}

\subsection{Uniform source, no dust}
Next, consider the case of a uniformly distributed source of \Lya\ in the cloud, with emissivity $w$. Physically, such scenario describes the case of gas cloud cooling via \Lya\ emission by hydrogen atoms, in the case of uniform density. The definition of the force multiplier factor needs to be changed to
\begin{equation}
\label{eqn:new-definition-MF}
M\ped{F}=\frac{F\ped{rad}}{\frac43\pi R^3w/c}.
\end{equation}

In this case, our analytical expression is
\begin{equation}
\label{eqn:MF-uniform}
M\ped{F}\ap{std}\approx\frac{12}{\pi^3}\sqrt{\frac23}\int_{-\infty}^{+\infty}\biggl(\Li_3\Bigl(e^{-\frac{\pi|\sigma(x)|}{\tau_0}}\Bigr)-\frac18\Li_3\Bigl(e^{-\frac{2\pi|\sigma(x)|}{\tau_0}}\Bigr)\biggr)\dd x,
\end{equation}
with a large-$x$ limit of
\begin{equation}
\label{eqn:MF-uniform-asymptotic}
M\ped{F}\ap{std}\approx0.51(\tau_0a_v)^{1/3},  
\end{equation}
in agreement with \citet{Lao:2020ptq}.
This result is also plotted in Fig.~\ref{fig:central-nodust-comparison}. The smaller $M\ped{F}$ value for a uniform source is the result of the shorter mean path the photons have to take in order to exit the cloud.

\subsection{Central source with dust}

We now consider the case of dusty media, in which \Lya\ photons can be absorbed by grains. The effect is quantified by the ratio\footnote{We use $\sigma\ped{d}=1.607\times 10^{-21}(D/D_\odot) \rm cm^2$ at the \Lya\ frequency, appropriate for a Milky Way, $R_V=3.1$ extinction curve \citep{weingartner2001}.}
\be
\epsilon\equiv\frac{\sigma\ped{d}}{\sigma_0}\approx 2.73\times10^{-8}\ T_4^{1/2}\left(\frac{D}{D_\odot}\right),
\ee
which appears in \cref{eqn:radtrasftausigma}; $D$ is the dust-to-gas mass ratio. In general, dust reduces the effectiveness of \Lya\ pressure. The impact of dust, however, heavily depends on the cloud optical depth. If $\tau_0$ is smaller than a certain critical value, $\tau_0^*$, dust absorption will not cause significant departures from the $\epsilon=0$ solution (\cref{eqn:MFstd-central-nodust}). On the other hand, for $\tau_0\gg\tau_0^*$, all radiation will ultimately be absorbed by dust, so $M\ped{F}$ is expected to reach a finite limit:
\be M\ped{F}(\infty,D)\equiv\lim_{\tau_0\to\infty}M\ped{F}(\tau_0,D).\ee 

In Appendix~\ref{sec:dust-details} we give a full mathematical derivation of $M\ped{F}(\infty,D)$. Here, let us show how its functional dependence on $a_v$ and $\epsilon$ can be obtained from simple arguments. Physically, $\tau_0^*$ corresponds to the optical depth of the cloud such that $\tau\ped{d}=n\ped{H}\sigma\ped{d}d=1$, where $d$ is the total distance travelled by photons. \Lya\ photons undergo a random walk in both frequency (changing at each scattering their frequency by a thermal width, $\Delta x=1$) and real space, with a mean free path $\ell=(n\ped{H}\sigma_0 H)^{-1}$ between collisions. 

After $N$ scatterings, the nondimensional r.m.s. photon frequency is $x=N^{1/2}$, with a travelled distance\footnote{We assume that scattering occurs in the line wings, where $ H(x)\propto x^{-2}$.},  
\be d=\sum_{i=0}^N \ell_i = \sum_{i=0}^N\frac1{n\ped{H}\sigma_0 H(i^{1/2})}=\frac{\sqrt\pi N^2}{2n\ped{H}\sigma_0a_v};\ee
by then, the photon will be at a r.m.s. distance from the center
\be r=\Bigg[\sum_{i=0}^N \ell_i^2\Bigg]^{1/2}=\sqrt{\frac{\pi}{3}}\frac{N^{3/2}}{n\ped{H}\sigma_0a_v}.
\label{eqn:rN32}\ee 
Further impose that $\tau\ped{d}=n\ped{H}\sigma\ped{d}d=1$ and $n\ped{H}\sigma_0 r=\tau_0^*$, to find
\begin{equation}
\tau_0^*=\frac{2^{3/4}\pi^{1/8}}{\sqrt{3}}a_v^{-1/4}\epsilon^{-3/4}=3.58\times10^6\ T_4^{-1/4}\biggl(\frac{D_\odot}{D}\biggr)^{3/4}.
\label{eqn:tau0-star}
\end{equation}
An equivalent derivation of \cref{eqn:tau0-star}, was shown by \citet{Hansen:2005pm}, from which we only differ by numerical factors.
This value of the optical depth will correspond to a \quotes{knee}, where the rising trend of $M_{\rm F}$ flattens and becomes independent of $\tau_0$. Note that this solution highlights the functional dependence of the force multiplier on temperature and dust-to-gas ratio. 

To help physical intuition we notice that for a solar metallicity gas at $T=100$ K, $M\ped{F}$ saturates at a column density $N\ped{H} = 6\times 10^{18}\, \rm cm^{-2}$. This value is much smaller than the typical $\approx 10^{20}\, \rm cm^{-2}$ column densities of galaxies and gas complexes within them, so estimating the saturation value is relevant, as it applies to most practical cases. 
This is computed as
\be M\ped{F}(\infty,D)=3.51(a_v\tau_0^*)^{1/3}=3.65\, \frac{a_v^{1/4}}{\epsilon^{1/4}}.\ee 

As anticipated, in Appendix~\ref{sec:dust-details} we show how the same conclusion can be derived from a more formal approach; in particular, the idea is to use the WKB approximation to put bounds on the solution of \cref{eqn:radtrasftausigma} when $\epsilon\ne0$. There, we obtain
\be
\label{eqn:MFstd-central-dust}
M\ped{F}(\infty, D) ={\cal N}\,\frac{a_v^{1/4}}{\epsilon^{1/4}}=11.5\, {\cal N}\ T_4^{-1/4}\biggl(\frac{D_\odot}{D}\biggr)^{1/4},
\ee
with an estimated order of unity pre-factor ${\cal N}=3.06$ (with accuracy below $10\%$) in excellent agreement with the physical derivation above. 

The same qualitative $M\ped{F}$ behaviour, with a knee followed by a flat trend for $\tau_0 > \tau_0^*$, was found with Monte-Carlo simulations by \citet{Kimm18}. We also nicely agree on the position of the knee, $\tau_0^*$. The asymptotic value of $M\ped{F}$, however, is about a factor of 2 larger than their result, reflecting our higher $M\ped{F}(\tau_0,D=0)$ curve, as explained before. Also the flattening reflects the physical process by which dust limits \Lya\ radiation pressure by absorbing UV photons, and re-emitting them in the infrared. Unless the column density is extremely high (see Sec. \ref{sec:UV}) re-emitted photons travel freely in the cloud, eventually escaping from it. 

\section{Implications for radiative feedback}
Radiative feedback, together with supernova-driven mechanical feedback, is an important regulator of galaxy evolution \citet{Ciardi05}. Among the various flavors of radiative feedback, here we are interested in assessing the relative importance of radiation pressure associated with three different physical mechanisms. 

We can identify at least three types of radiation forces produced by (a) hydrogen photoionization, $F_i$; (b) UV light extinction by dust, $F_{\rm UV}$; (c) Ly$\alpha$ scattering, the focus of this paper, $F_\alpha$. Our aim is to clarify the different dependencies of these three forces, and quantify their relative contribution as a function of the key parameters, the gas column density, $N\ped{H}$, and metallicity $Z$. 

We will make the assumption that the dust-to-gas ratio scales linearly with $Z$, and normalize such relation to the Milky Way values: $D/D_\odot = Z/Z_\odot$, and we take $D_\odot = 1/162$ \citep{Remy14}, and $Z_\odot=0.0142$ \citep{asplund2009}. Hence, we can simply write $D=f_d Z = 0.43\, Z$, by suitably defining the dust-to-metal ratio $f_d=D_\odot/Z_\odot$. 

We now proceed to write explicit expressions for the three forces. We note that, as the ionizing photon rate, $Q$, and the specific UV continuum luminosity, $L_{\lambda, \rm UV}$, entering the expression for the three forces, both depend linearly on the star formation rate, the comparison among them is independent of such quantity. In the following, thus, we will give these quantities per star formation rate in units of $M_\odot \rm yr^{-1}$.

\subsection{Photoionization}\label{sec:photo}
The force associated with the photoionization of an hydrogen atom by a Lyman continuum (LyC) photon with energy $h\nu > 13.6\, {\rm eV} \equiv 1\, {\rm Ryd}$ is
\be\label{eq:Fi}
F_i = \frac{L_i}{c}(1-e^{-\tau_{\rm HI}}).
\ee
In the previous expression, $L_i = Q \langle h \nu\rangle_i$ is the total LyC luminosity of the source; $\langle h \nu\rangle_i$ is the cross section-weighted ionizing photon energy in the source spectrum. The \HI optical depth, often cast in terms of the escape fraction of ionizing photons $f_{\rm esc} = e^{- \tau_{\rm HI}}$, can be written\footnote{We neglect the small extra opacity to ionizing photons due to dust. We also assume that the gas is mostly neutral, i.e. $N\ped{H} \simeq N_{\rm HI}$, as the depth of the \HII region is typically much smaller than the size of the system in many astrophysical situations.} as $\tau_{\rm HI}\simeq \sigma_{\rm L} N_{\rm HI}  ({\rm Ryd}/\langle h \nu\rangle_i)^3$, where $\sigma_L=6.3\times 10^{-18} \rm cm^2$.

The ionizing photon rate $Q$ (per unit star formation rate) depends on the initial mass function (IMF) and on the metallicity of the stellar (the calculation can be extended to non-stellar sources in a straightforward manner) sources. We assume a standard $1-100 M_\odot$ Salpeter IMF and continuous star formation. Then, a convenient expression, that we adopt here, is \citep{Schaerer03}: 
\be\label{eq:Q}
\log Q = -0.0029 (\log Z + 9.0)^{5/2} + 53.81. 
\ee

\subsection{UV extinction by dust}\label{sec:UV}
Non-ionizing UV photons can transfer momentum to dust grains when they are absorbed/scattered. As dust is usually tightly coupled to the gas via hydrodynamical and Coulomb drag forces, 
the momentum is ultimately transferred to the gas. The force associated with this dust-mediated radiation pressure is 
\be\label{eq:FUV}
F_{\rm UV} = \frac{L_{\rm UV}}{c}[(1-e^{-\tau_{\rm d}}) + f\ped{IR}],
\ee
where $f\ped{IR} \simeq \tau\ped{IR}$ \citep{krumholz2012} allows from dust re-emitted IR photons to contribute to the pressure if the cloud IR optical depth $\tau\ped{IR}$ is sufficiently high. We write this term following \citet{pallottini2017}:
\be
\tau\ped{IR} = 1.79\times 10^{-24} \left(\frac{N\ped{H}}{\rm cm^{-2}}\right)\left(\frac{D}{D_\odot}\right)\left(\frac{T\ped{d}}{100\, \rm K}\right)^2,
\ee
where $T\ped{d}$ is the dust temperature.
Consider the UV luminosity in the Habing band ($6 < h\nu/{\rm eV} < 13.6$), whose width is therefore $\Delta \lambda = 1155\,$\AA. For the adopted IMF, the specific luminosity per unit star formation rate, $L_{\lambda,\rm UV}$ is obtained using \textsc{STARBURST99}. It turns out that the dependence on $Z$ is very weak and therefore we neglect it; also, $L_{\lambda,\rm UV} \simeq 3.5 \times 10^{40} {\rm erg s^{-1}}$ \AA$^{-1} \approx$ const. in the band. Finally, $L_{\rm UV} = L_{\lambda,\rm UV} \Delta \lambda$.

The average dust optical depth in the Habing band is $\tau_d = \langle\sigma_d\rangle N_{\rm H} = 1.91 \times 10^{-21} (D/D_\odot) N_{\rm H}$  
appropriate for a Milky Way, $R_V=3.1$ extinction curve \citep{weingartner2001}.

\subsection{\Lya\ scattering}\label{sec:Lya}
From the previous results, it follows that
\be\label{eq:Fa}
F_{\alpha} = M\ped{F} \frac{L_{\alpha}}{c},
\ee
where 
\be
L_{\alpha} = \frac{2}{3} Q (1 - e^{-\tau_{\rm HI}}) h \nu_\alpha,
\ee
and 
\be
M\ped{F}=\min\Big[M\ped{F}(\tau_0,D=0),M\ped{F}(\infty,D)\Big].
\ee
The two expressions for $M\ped{F}$ are given by \cref{eqn:MFstd-central-nodust} and \cref{eqn:MFstd-central-dust}, respectively.

\subsection{Comparison}\label{sec:comp}
Armed with the above results, we can now quantitatively compare the three radiative forces. We reiterate that the following conclusions \textit{do not depend on the star formation rate} of the galaxy, due to the common linear dependence of the forces on such quantity. However, our findings depend on the assumed IMF, which we take here to be a standard Salpeter one, as described in Sec. \ref{sec:photo}.

We start by comparing photoionization and dust-mediated UV radiation pressures, by analysing the ratio $F_i/F_{\rm UV}$ as a function of $N\ped{H}$ and $Z$. This is done using the expressions given by eq. \ref{eq:Fi}  and \ref{eq:Fa}, and the results are shown in the left panel of Fig. \ref{Fig02}. Photoionization dominates in the low $N\ped{H}$, low $Z$ regime, where the decreasing dust opacity makes UV radiation pressure less efficient. For a typical molecular cloud in the Milky Way, characterized by $N\ped{H} \simeq 10^{22} \rm cm^{-2}$ and $Z \simeq Z_\odot$, we find that radiation pressure on dust largely  
dominates over photoionization. However, in a more metal-poor environment ($Z \simlt 10^{-2} Z_\odot$), typical early galaxies photoionization takes over. 

However, an inspection of the right panel shows that the above forces are most often overwhelmed by \Lya\ pressure which therefore controls the dynamics of the cloud. There we show the ratio $F_\alpha/(F_i+F_{\rm UV})$ in the same $N\ped{H}-Z$ plane. \Lya\ pressure dominates the entire parameter range, with the only exception of a small region (upper-right) corner of extremely high column densities and super-solar metallicities. The ratio between \Lya\ and the other two forces is always $\simgt 10$, and reaches values as high as $\approx 300$ for low $Z$ and high $N\ped{H}$ (red triangle in the panel).
Thus, we conclude that \Lya\ has important dynamical effects, at least in the idealised situation of a uniform, spherical and static cloud we are dealing with. 

As a final remark, it is useful to note that the comparison made above assumes that the various feedbacks act co-spatially on the gas surrounding the sources. In reality, ionizing photons transfer momentum on the relatively small scale of \HII regions, whereas \Lya\ pressure works on larger volumes, whose extent depends on the dust opacity, and hence metallicity of the cloud. 
\section{Summary and discussion}\label{sec:summary}
We have developed a model describing the radiation pressure due to \Lya\ line photons, and obtained analytical expressions for the force-multiplier, $M\ped{F}(N\ped{H}, Z)=F_\alpha/(L_\alpha/c)$ as a function of gas column density, $N\ped{H}$, and metallicity, $Z$, for both dust-free and dusty media. The key result is that $M\ped{F}$ can be expressed as \be
M\ped{F}=\min\Big[M\ped{F}(\tau_0,D=0),M\ped{F}(\infty,D)\Big];
\ee
where two expressions for $M\ped{F}$ are given by \cref{eqn:MFstd-central-nodust} and \cref{eqn:MFstd-central-dust}, respectively. This solution has been compared with the numerical ones by \citet{Kimm18} finding a good agreement in terms of the general trend apart from an essentially constant, order unity factor of unknown origin. Numerical approaches to the problem  might include additional physical effects and deal with less idealized scenarios. Yet, a good analytical understanding of the underlying physics is helpful, and analytical estimates may be used to speed-up or test simulations.

As our analytical solutions include dust, it is possible to determine the relative importance of \Lya\ pressure compared to two other physical mechanisms also transferring momentum from radiation to the gas: photoionization and dust-mediated UV radiation pressure. We find that, in static media, \Lya\ pressure dominates over both photoionization and dust-mediated UV radiation pressure in a very wide parameter range ($16 < \log N\ped{H} < 23; -4 < \log [Z/Z_\odot] < 0$). For example, it overwhelms the other two forces by $\simgt 10$ (300) times in standard (low-$Z$) star-forming clouds.

In agreement with several previous studies, we conclude conclude that \Lya\ pressure plays a dominant role in the initial acceleration of the gas around luminous sources, and should be properly implemented in galaxy formation, evolution and outflow models and simulations. Several consequences of the present results deserve further attention. Among these, \Lya\ radiation feedback might act on star forming molecular clouds by pushing gas away from star forming clumps, thereby reducing the star formation rate and/or preventing very strong bursts. Also, supernovae (SNe) exploding in this pre-conditioned environment will be less effective in collecting material and eject it, with the consequence that the outflow mass-loading factor (outflow rate per unit star formation rate) could be significantly reduced \citet{Kimm18}. 

It is worth noting that the momentum rate per unit star formation rate of \Lya\ radiation is comparable to that injected by SNe. Note that such ratio is independent of star formation rate. Using eq. \ref{eq:Fa}, and taking $Q$ from  eq. \ref{eq:Q}, we find $F_\alpha = 2.34\times 10^{32} M_{\rm F} = 2.34 \times 10^{34} {\rm g\, cm\, s^{-2}}$ (per unit SFR in $M_\odot {\rm yr}^{-1}$, and for for $Z=Z_\odot$). For an order of magnitude estimate, the analogous quantity for SNe is $F_{\rm SN} = 1.4\times 10^{33}{\rm g\, cm\, s^{-2}}$, having assumed that, consistently with the IMF adopted in Sec. \ref{sec:photo}, one SN forms every $135\, M_\odot$ of stars, and ejects 10 $M_\odot$ of material at a terminal speed of $3000\, \rm km\, s^{-1}$. Thus, at face value, although SNe and \Lya\ feedback operate at different times, momentum injection by \Lya\ radiation pressure exceeds the one by SNe by a factor of $\simgt 10$.

We can also compare the \Lya\ force with the gravity force, $F_g$, in a molecular cloud with a given turbulent velocity dispersion $\sigma_{\rm kms} = \sigma/({\rm km\, s}^{-1})$. As $F_g = 4 \pi R^2 p$, where $p$ ($R$) is the cloud pressure (radius), and recalling that $R = \sigma^2/2\sqrt{Gp}$ \citep{Sommovigo20}, we find $F_g = \pi \sigma^4/G$. The SFR in the cloud can be written as 
\begin{equation}
    \psi = \epsilon_{\rm ff} \frac{M}{t_{\rm ff}} = \frac{1}{2}\epsilon_{\rm ff}\frac{\sigma^3}{G}
    = 10^{-5.9} \sigma_{\rm kms}^3 {\msun}{\rm yr^{-1}},
    \label{sfr}
\end{equation}
where $t_{\rm ff}$ is the cloud free-fall time, and $\epsilon_{\rm ff}\approx 0.01$ is the amount of gas converted in stars within $t_{\rm ff}$. With these formulae and the previous estimate for $F_\alpha$, we find that \Lya\ radiation pressure exceeds the gravity force if the condition $M_{\rm F} > 20 \sigma_{\rm kms}$ is satisfied. As for $Z=Z_\odot$, $M_{\rm F}  \approx 100$, we see that in GMCs, for which $\sigma_{\rm kms} = 1-5$ \citep{Ballesteros11}, gravity plays a subdominant role.

To obtain an analytical solution in closed form, various approximations have been made: for example, no recoils, uniform density and static gas. As mentioned before, the effect of recoils may be responsible for at least part of the discrepancy with \citet{Kimm18}. Regarding the uniform density, we do not expect significant deviations if spherical symmetry is maintained (e.g.~in the no-dust case, the scaling is still expected to be 1/3, \citet{Lao:2020ptq}), while asymmetries or ``cavities'' in the medium (\citet{Behrens:2014bua}) might have a large impact, though difficult to estimate. The static approximation, however, is particularly critical because it is guaranteed to be, to some extent, violated due to the effect of the \Lya\ pressure itself. The results above indicate that \Lya\ radiation pressure is certainly the dominant source of momentum for the gas, both with respect to other radiative processes and SNe, when the source is turned on and the gas is presumably close to static. Hence, there is little doubt that in the initial acceleration phase of the surrounding material the contribution of \Lya\ photons cannot be neglected. However, as the gas is set in a bulk, outflowing motion away from the luminous source (e.g. a stellar cluster or an accreting black hole) $F_\alpha$ is likely to decrease as the photons are shifted out of the line resonance. 


\begin{figure}
\includegraphics[width=0.48\textwidth]{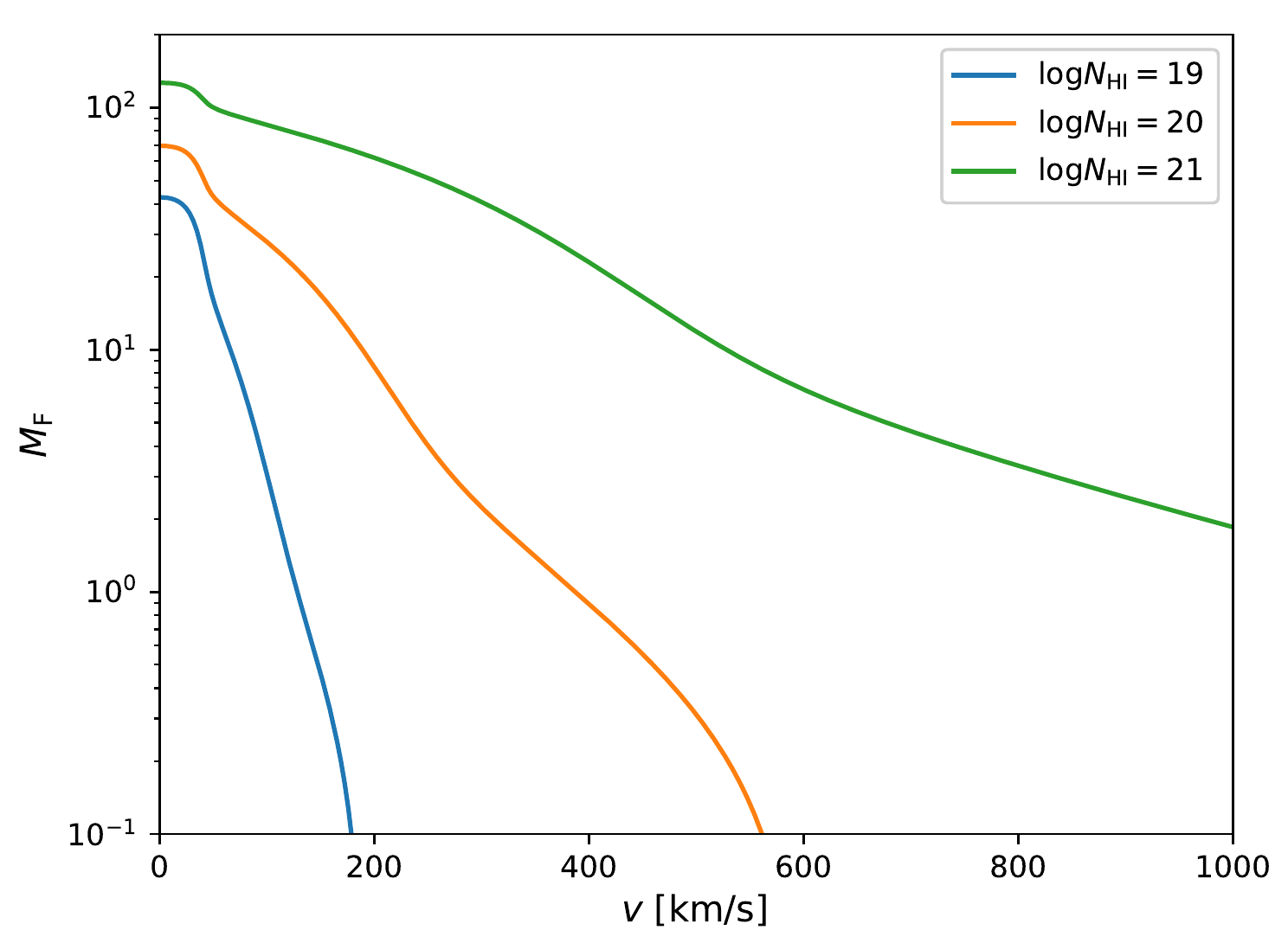}
\caption{Dependence of the force multiplier, $M_{\rm F}$, on the gas bulk velocity, $v$, for three values of the gas column density shown in the label. All curves refer to a dust-free medium ($D=0$) and assume a gas temperature $T_4=1$. Compare to Fig.~3 of \citet{Dijkstra09}.
\label{fig:MF(v)} }
\end{figure}

To approximately quantify this effect in the dust-free case, instead of a gas moving at a bulk velocity $v=x_{\rm s}v_{\rm th}$, we redshift the injected photons by an amount $(v/c)\,\nu_0$.  By applying \cref{eqn:M_F-shifted} it is then easy to compute $M_{\rm F}(v, N_{\rm H})$. The results are shown for different column densities (or, equivalently $\tau_0$) in Fig.~\ref{fig:central-nodust-comparison} (orange thin curves), and in Fig.~\ref{fig:MF(v)} as a function of velocity. First, we note (Fig.~\ref{fig:central-nodust-comparison}) that for $\tau_0 \to \infty$ the effect of the shift gets washed out, because of diffusion in photon frequency. \Lya\ radiation pressure decreases, at fixed $N_{\rm H}$, with velocity, and the velocity that leads to a significant reduction of $M\ped{F}$ depends on $N\ped{H}$. For an order of magnitude estimate, we reduce to a single-scattering scenario ($M\ped{F}=1$) when the wing optical depth is of order unity:
\be
\frac{\tau_0 a_v}{\sqrt\pi x\ped{s}}=1\implies v=x\ped{s}v\ped{th}=\SI{500}{km/s}\,\sqrt{\frac{N\ped{HI}}{\SI{e+20}{cm^{-2}}}}.
\ee
The drop is less pronounced for high column densities where frequency diffusion effects become more important. For example a factor 3 reduction is achieved for $v \simeq (50, 140, 300)\, {\rm km\, s}^{-1}$ for $\log N_{\rm H}= (19,20,21)$. An equivalent plot, but in the case of a spherical shell, was presented by \citet{Dijkstra09} (see their Fig. 3) with a numerical approach, and is in reasonable agreement with our analytical computation. We conclude that a bulk velocity does not  induce a dramatic reduction in the \Lya\ radiation pressure if the \HI column density is sufficiently large. It is worth noting that, instead, intense velocity gradients may significantly enhance escape and thus, presumably, reduce $M\ped{F}$ (\citet{Seon:2020}).

Another type of motion that the idealized scenario does not take into account is turbulence. While a proper treatment undoubtedly requires a numerical approach, an idea may be mimicking it using a larger ``effective'' temperature $T^{\rm turb}$ of the gas. As $M\ped{F}\propto T^{-1/6}$, this effect may reduce $M\ped{F}$ as
\be
\frac{M\ped{F}^{\rm turb}}{M\ped{F}}=\biggl(\frac{T}{T^{\rm turb}}\biggr)^{1/6}=\biggl(1+0.006\,\frac{\sigma\ped{kms}^2}{T_4}\biggr)^{-1/6}.
\ee
This reduction should thus not change the overall importance of \Lya\ pressure in most cases.

The effects of \Lya\ radiation pressure might become more important in removing the gas from the star forming site \textit{before} SNe can go off. As we have seen, the ratio $F_\alpha/(F_i+F_{\rm UV}) >10$ for any metallicity (Fig. \ref{Fig02}). As molecular clouds are characterized by large neutral gas column densities we can ignore to a first approximation the reduction effects as the gas accelerates up to their escape velocity ($v_{\rm e} \simeq 20\, {\rm km\, s}^{-1}$).  Thus, for a typical star formation rate of a molecular cloud $\psi \approx 10^{-3} M_\odot \rm yr^{-1}$, before the onset of SNe after $\Delta t \approx 3$ Myr, \Lya\ pressure has already cleared about $F_\alpha \psi \Delta t/v_{\rm e} \simeq 10^5 M_\odot$ of gas, i.e. a sizeable (if not all) mass fraction of the cloud has been accelerated to velocities exceeding the cloud escape speed. When the first SN explosions take place, they will find a much rarefied environment. Additional discussion on these combined feedback effects is given in \citet{Kimm19, Abe18}.

 Another competing effect if associated with the gas photoheating inside the \HII region created by the massive stars. The \HII region makes a transition to a D-type after approximately a recombination time $(\alpha_B n)^{-1} = 1.22$ kyr, where $\alpha_B =2.6\times 10^{-13} {\rm cm}^3 \rm s^{-1}$ is the Case-B recombination coefficient at temperature $T \approx 10^4$ K, and $n \approx 100\, \cc$ is the mean gas density. Although this timescale is short and comparable to the dynamical one imposed by the \Lya\ force, the \HII region expansion initially occurs with a velocity comparable to the sound speed of the ionized gas, $\simeq 10\, {\rm km\, s}^{-1}$, and decreases as $(r/R_S)^{-3/4}$ beyond the initial Str\"omgren radius, $R_{\rm S}$. Thus, the expansion velocity is significantly smaller than the one imposed by \Lya\ radiation pressure.

Although the above dimensional calculations clearly ask for more detailed studies, they indicate that the impact of \Lya\ radiation pressure cannot be anymore neglected as simulations and models struggle for a more complete physical description of the ISM and its dynamics. This is even more urgent in early galaxies models, where the lower dust content strongly enhances the role of \Lya\ radiation pressure.

\section*{acknowledgements}
We are indebted to the referee, A. Smith, for a number of useful suggestions and points. AF acknowledges support from the ERC Advanced Grant INTERSTELLAR H2020/740120. Any dissemination of results must indicate that it reflects only the author’s view and that the Commission is not responsible for any use that may be made of the information it contains. Support from the Carl Friedrich von Siemens-Forschungspreis der Alexander von Humboldt-Stiftung Research Award is kindly acknowledged.


\appendix
\section{Analytical computation of radiation pressure}
\label{app:computations}

Start from the stationary radiative transfer equation,
\be \vec{\ver n\cdot\nabla}I(\vec r, \ver n, \nu)=-\alpha\ped{abs}I(\mathbf{r},\mathbf{\hat n},\nu)+j\ped{source}+j\ped{scatt}.\ee 
Given the spherical symmetry of the problem, $I$ will only be a function of $r$ and the angle between $\vec r$ and $\ver n$, which is encoded for convenience in $\mu=\vec{\hat n\cdot\hat r}$. Writing explicitly the gradient of $I(r,\mu,\nu)$ in spherical coordinates, we obtain
\begin{equation}
\mu\frac{\partial I}{\partial r}+\frac{1-\mu^2}{r}\frac{\partial I}{\partial \mu}=-\alpha\ped{abs}I+j\ped{source}+j\ped{scatt},
\label{eqn:radtrasf}
\end{equation}
Let us define the moments of the specific intensity as follows:
\be J=\frac{1}{2}\int_{-1}^1I\dd\mu,\qquad\tilde H=\frac{1}{2}\int_{-1}^1I\mu\dd\mu,\qquad K=\frac{1}{2}\int_{-1}^1I\mu^2\dd\mu,\ee 
and similarly for the ``source'' and ``scattering'' terms (the tilde on $\tilde H$ is to avoid confusion with the Voigt profile).

In the limit of large optical depth, we employ the Eddington approximation: $I=a(r,\nu)+b(r,\nu)\mu$, with $b\ll a$, implying $J=3K$. Taking the integral over $\dd\mu$ of \cref{eqn:radtrasf} and of \cref{eqn:radtrasf} multiplied by $\mu$, we find
\begin{gather}
\label{eqn:integral-mu}
    \frac{\partial \tilde H}{\partial r}+\frac{2\tilde H}r=-\alpha\ped{abs}J+J\ped{source}+J\ped{scatt},\\
\label{eqn:mu-then-integral-mu}
    \frac13\frac{\partial J}{\partial r}=-\alpha\ped{abs}\tilde H+\tilde H\ped{source}+\tilde H\ped{scatt}.
\end{gather}

We will assume the external sources to be isotropic, therefore $\tilde H\ped{source}=0$. It is less obvious that $\tilde H\ped{scatt}=0$, but this is the case, as we will now prove. The angular distribution of \Lya\ photons scattered by hydrogen atoms is a superposition of isotropic and dipole distributions. Given an incident intensity of the form $a+b\cos\theta$, as per Eddington approximation, an isotropic scattering function gives an isotropic light redistribution:
\be \frac1{4\pi}\int_0^\pi\sin\tilde\theta\dd\tilde\theta\int_0^{2\pi}\dd\tilde\varphi\, (a+b\cos\tilde\theta)=a.\ee 
The computation for the dipole scattering is less trivial, but again straightforward:
\be \frac1{4\pi}\int_0^\pi\sin\tilde\theta\dd\tilde\theta\int_0^{2\pi}\dd\tilde\varphi\, (a+b\cos\tilde\theta)\frac34\Bigl(1+\cos^2\chi\Bigr)=a,\ee 
where
\be \cos\chi=\cos\theta\cos\tilde\theta+\sin\theta\sin\tilde\theta\cos(\varphi-\tilde\varphi)\ee 
is the cosine of the angle between the final and initial directions, $(\theta,\varphi)$ and $(\tilde\theta,\tilde\varphi)$. Therefore, no matter what the ratio between isotropic and dipole emission, the scattered light will always be isotropic according to the Eddington approximation, i.e., $\tilde H\ped{scatt}=0$.

Combining \cref{eqn:integral-mu} and \cref{eqn:mu-then-integral-mu}, we get
\be -\frac{1}{3\alpha\ped{abs}}\biggl(\frac{\partial^2J}{\partial r^2}+\frac{2}{r}\frac{\partial J}{\partial r}\biggr)=-\alpha\ped{abs}J+J\ped{source}+J\ped{scatt}.\ee 
It is convenient to use the optical depth as a variable instead of the physical radius. By defining the optical depth at the line centre as $\tau=n\ped{H}\sigma_0r$, we get
\be \frac{\partial^2J}{\partial \tau^2}+\frac{2}{\tau}\frac{\partial J}{\partial \tau}=\frac{3\alpha\ped{abs}^2}{n\ped{H}^2\sigma_0^2}J-\frac{3\alpha\ped{abs}}{n\ped{H}^2\sigma_0^2}J\ped{source}-\frac{3\alpha\ped{abs}}{n\ped{H}^2\sigma_0^2}J\ped{scatt}.\ee 

In an optically thick medium, the majority of the photons are in the wings of the Voigt profile. This happens because the mean free path for core photons is so short that they keep scattering until they exit the core region to enter the wing, where the cross section is lower and they can stay for a long time. This effect can be encoded in the ``Fokker-Planck'' approximation (see \citet{Dijkstra17} for a derivation):
\be J\ped{scatt}=n\ped{H}\sigma_0 H(x)J+n\ped{H}\sigma_0\frac{1}{2}\frac{\partial}{\partial x}\biggl( H(x)\frac{\partial J}{\partial x}\biggr).\ee 

Following \citet{Harrington73}, we define
\be \sigma(x)=\int_0^x\frac{1}{\sqrt{3/2} H(y)}\dd y,\ee 
so that
\be \frac{1}{2}\frac{\partial}{\partial x}\biggl( H\frac{\partial J}{\partial x}\biggr)=\frac{1}{3 H}\frac{\partial^2J}{\partial\sigma^2}.\ee 
The radiative transfer equation then becomes
\begin{equation}
\frac{\partial^2J}{\partial \tau^2}+\frac{2}{\tau}\frac{\partial J}{\partial \tau}+\frac{\partial^2J}{\partial\sigma^2}=3 H\frac{\sigma\ped{dust}}{\sigma_0}J-\frac{3 H}{n\ped{H}\sigma_0}J\ped{source},
\label{eqn:radtrasftausigma-app}
\end{equation}
where we only kept the lowest order in $\sigma\ped{dust}/\sigma_0$ in the coefficient of each term.

\subsection{Central source, no dust}
Let 
\be J\ped{source}=\frac{W}{4\pi r\ped{s}^2}\delta(r-r\ped{s})\delta(\nu-\nu_0)=\frac{Wn\ped{H}^3\sigma_0^3}{4\pi \tau\ped{s}^2\Delta\nu\ped{0}}\delta(\tau-\tau\ped{s})\delta(\sigma)\frac{1}{\sqrt{3/2} H}.\ee 
We will then let $r\ped{s}\to0$ to obtain the limit of a central point-source. Substituting into~\cref{eqn:radtrasftausigma-app} and writing it in nondimensional form, we get
\begin{equation}
\frac{\partial^2f}{\partial \tau^2}+\frac{2}{\tau}\frac{\partial f}{\partial \tau}+\frac{\partial^2f}{\partial\sigma^2}=-\frac{\delta(\tau-\tau\ped{s})\delta(\sigma)}{\tau\ped{s}^2},
\label{eqn:eqnc}
\end{equation}
where $f=J/J_0$ and $J_0={\sqrt{6}Wn\ped{H}^2\sigma_0^2}/{4\pi\Delta\nu\ped{0}}$.

For the boundary condition in the frequency domain, we require
\be \lim_{\sigma\to\pm\infty}f(\tau,\sigma)=0,\ee 
that is, the specific intensity will remain concentrated in a neighbourhood of the line center. In the space domain, as anticipated in the main text, we will use two different boundary conditions, in order to show that they do not play a crucial role:
\be \biggl(\frac{\partial f}{\partial\tau}\biggr)_{\tau=\tau_0}=\begin{cases}
-\frac{3}{2} H(\sigma) f(\tau_0,\sigma) & \text{standard,}\\
0 & \text{alternative.}
\end{cases}\ee 
\citet{Harrington73} and \citet{Dijkstra06} use the ``standard'' one, which comes from imposing the Eddington approximation on the boundary (one integrates over $\mu$ from $-1$ to 0 and gets $J=2\tilde H$) and combining it with \cref{eqn:mu-then-integral-mu}. The reason we are investigating a different boundary condition (the flux-free is just a simple example of such) is that it could be argued that, at the edge of the cloud, the Eddington approximation fails, and thus the $-3/2$ factor is not a credible value. Other authors also discussed this issue, for example \citet{Lao:2020ptq} (see their Appendix A) proposed a factor of $\sqrt3$ as a result of the two-stream approximation.

Separating variables, we look for a solution of the form
\be f(\tau,\sigma)=\sum_{n=1}^{\infty}T_n(\tau)S_n(\sigma),\ee 
with
\be \frac{\partial^2T_n}{\partial\tau^2}+\frac{2}{\tau}\frac{\partial T_n}{\partial\tau}+\lambda_n^2T_n=0\implies T_n=A\frac{\sin(\lambda_n\tau)}{\lambda_n\tau}+B\frac{\cos(\lambda_n\tau)}{\lambda_n\tau}.\ee 
In order to avoid the specific intensity to diverge at \emph{all} frequencies near the origin, we require $B=0$. The eigenvalues $\lambda_n$ are determined substituting into the boundary conditions:
\be \tau_0\lambda_n\cot(\lambda_n\tau_0)-1=\begin{cases}
-\frac{3}{2}\tau_0 H & \text{standard}\\
0 & \text{alternative}
\end{cases}\ee 
meaning
\be \lambda_n\approx\begin{cases}
\frac{n\pi}{\tau_0}\bigl(1-\frac{2}{3\tau_0 H}\bigr) & \text{standard ($\tau_0 H\gg1$)}\\
\frac{n\pi}{\tau_0}-\frac{\pi}{2\tau_0}-\frac{1}{(n-1/2)\pi\tau_0} & \text{alternative.}
\end{cases}\ee 

Then, we normalize $T_n$ as
\be \int_0^{\tau_0}4\pi\tau^2T_n^2\dd\tau=1\implies T_n=\frac{\sin(\lambda_n\tau)}{\tau\sqrt{2\pi\tau_0}},\ee 
so that $f(\tau,\sigma)=\sum_{n=1}^\infty \frac{\sin(\lambda_n\tau)}{\tau\sqrt{2\pi\tau_0}}S_n(\sigma)$. Substituting into \cref{eqn:eqnc} and projecting on a specific $n$,
\be \begin{split}
    \frac{\partial^2S_n}{\partial\sigma^2}-\lambda_n^2S_n&=-\int_0^{\tau_0}4\pi\tau^2\frac{\delta(\tau-\tau\ped{s})\delta(\sigma)}{\tau\ped{s}^2}\frac{\sin(\lambda_n\tau)}{\tau\sqrt{2\pi\tau_0}}\dd\tau\\
    &=-\frac{4\pi\sin(\lambda_n\tau\ped{s})}{\tau\ped{s}\sqrt{2\pi\tau_0}}\delta(\sigma),
\end{split}\ee 
which gives
\be S_n=\sqrt{\frac{2\pi}{\tau_0}}\frac{\sin(\lambda_n\tau\ped{s})}{\lambda_n\tau\ped{s}}e^{-\lambda_n|\sigma|}\xrightarrow{\tau\ped{s}\to0}\sqrt{\frac{2\pi}{\tau_0}}e^{-\lambda_n|\sigma|}\ee 
and
\be f(\tau,\sigma)=\sum_{n=1}^\infty\frac{\sin(\lambda_n\tau)}{\tau_0\tau}e^{-\lambda_n|\sigma|}.\ee 

Substituting directly into \cref{eqn:Frad}, we get
\be M\ped{F}\ap{std}\approx\frac{4}{\pi}\sqrt{\frac{2}{3}}\int_{-\infty}^{+\infty}\artanh\bigl(e^{-\pi|\sigma(x)|/\tau_0}\bigr)\dd x,\ee 
and
\begin{equation}
\label{eqn:MF-central-alternative}
\begin{split}
M\ped{F}\ap{alt}\approx{}&\frac{4}{\pi}\sqrt{\frac{2}{3}}\int_{-\infty}^{+\infty}\artanh\Bigl(e^{-\frac{\pi|\sigma(x)|}{2\tau_0}}\Bigr)\dd x-{}\\
&-\int_{-\infty}^{+\infty}\frac{\dd x}{\sqrt{6}\cosh\bigl(\frac{\pi|\sigma(x)|}{2\tau_0}\bigr)},\end{split}
\end{equation}
where the following series have been used:
\begin{align*}
\sum_{n=0}^\infty\frac{a^{-(2n+1)}}{2n+1}&=\artanh(1/a),\\
\sum_{n=1}^\infty\frac{(-1)^n}{a^n}&=-\frac1{1+a},\\
\sum_{n=1}^\infty\frac{a^{-n-1/2}}{n-1/2}&=2\artanh(a^{-1/2}).
\end{align*}

To get the asymptotic limit at large $\tau_0$, we sustitute the large-$x$ limit of $\sigma(x)$ and perform the integral over $\dd x$ before resumming the series. The result is
\be M\ped{F}\ap{std}\sim\frac{4(2^{4/3}-1)\Gamma(4/3)\zeta(4/3)}{\pi^{3/2}}(\tau_0a_v)^{1/3}\approx3.51\times(\tau_0a_v)^{1/3}\ee 
and
\be \begin{split}
M\ped{F}\ap{alt}\sim{}&\frac{2\Gamma(4/3)}{\pi^{1/2}}\biggl(\frac{2^{4/3}(2^{4/3}-1)\zeta(4/3)}\pi+\zeta\biggl(\frac13,\frac34\biggr)-\zeta\biggl(\frac13,\frac14\biggr)\biggr)\times{}\\
&\times(\tau_0a_v)^{1/3}\approx{}\\
\approx{}&3.43\times(\tau_0a_v)^{1/3},
\end{split}\ee 
where the integral $\int_0^\infty e^{-x^3}\dd x=\Gamma(4/3)$, and the definition of the Hurwitz zeta function have been used.

\subsubsection{Source emitting at a different frequency}

If
\[J\ped{source}=\frac{W}{4\pi r\ped{s}^2}\delta(r-r\ped{s})\delta(\nu-\nu\ped{s}),\]
with $\nu\ped{s}\ne\nu_0$, the only change to \cref{eqn:eqnc} is
\[\delta(\sigma)\to\delta(\sigma-\sigma(x\ped{s})),\]
with $x\ped{s}=\nu\ped{s}/\Delta\nu_0$. Therefore, the new formula for $M\ped{F}$ is simply
\be M\ped{F}\ap{std}\approx\frac{4}{\pi}\sqrt{\frac{2}{3}}\int_{-\infty}^{+\infty}\artanh\bigl(e^{-\pi|\sigma(x)-\sigma(x\ped{s})|/\tau_0}\bigr)\dd x.
\label{eqn:M_F-shifted}\ee
This case is relevant to compute \Lya\ radiation effects on a medium moving at a velocity $v$, which receives photons at a frequency different from $\nu_0$ due to Doppler effects. It has been applied to obtain the orange thin curves in Fig. \ref{fig:central-nodust-comparison}.

\subsection{Uniform source, no dust}

Let
\be J\ped{source}=w\delta(\nu-\nu_0)=\frac{w}{\Delta\nu_0\sqrt{3/2} H}\delta(\sigma).\ee 
Substituting into~\cref{eqn:radtrasftausigma-app} and writing it in nondimensional form, we get
\be \frac{\partial^2f}{\partial \tau^2}+\frac{2}{\tau}\frac{\partial f}{\partial \tau}+\frac{\partial^2f}{\partial\sigma^2}=-\delta(\sigma),\ee where $f=J/J_0$ and $J_0=\frac{\sqrt{6}w}{n\ped{H}\sigma_0\Delta\nu_0}$.
With the same approach as before, we find
\be f(\tau,\sigma)=\sum_{n=1}^\infty(-1)^{n-1}\frac{\sin(\lambda_n\tau)}{\lambda_n\tau}\frac{e^{-\lambda_n|\sigma|}}{\lambda_n}.\ee 

Making use of the definition \cref{eqn:new-definition-MF}, the final formula for the multiplication factor is\footnote{The polylogarithm of order 3 is defined as $\Li_3(z)=\sum_{n=1}^\infty\frac{z^k}{k^3}$.}
\be M\ped{F}\ap{std}\approx\frac{12}{\pi^3}\sqrt{\frac23}\int_{-\infty}^{+\infty}\biggl(\Li_3\Bigl(e^{-\frac{\pi|\sigma(x)|}{\tau_0}}\Bigr)-\frac18\Li_3\Bigl(e^{-\frac{2\pi|\sigma(x)|}{\tau_0}}\Bigr)\biggr)\dd x.\ee 
The expression stemming from alternative boundary conditions involves the Lerch transcendent function and we will not report it here. As before, we can resum the series in the large-$x$ limit to get
\be M\ped{F}\ap{std}\sim\frac{3\Gamma(4/3)}{\pi^{7/2}}(2^{10/3}-1)\zeta(10/3)(\tau_0a_v)^{1/3}\approx0.51(\tau_0a_v)^{1/3}.\ee 

\subsection{Central source with dust}
\label{sec:dust-details}

The only change to the radiative transfer equation \cref{eqn:radtrasftausigma-app} when $\sigma\ped{dust}\ne0$ is the addition of the term $3 H\sigma\ped{dust}/\sigma_0$. Variables can still be separated and $T_n=\frac{\sin(\lambda_n\tau)}{\tau\sqrt{2\pi\tau_0}}$ as before. Adopting standard boundary conditions, we can conveniently rewrite the formula for $M\ped{F}$ as
\be M\ped{F}=\frac4{\sqrt{2\pi\tau_0}}\sum_{n=0}^\infty\int_{-\infty}^{+\infty}\frac{S_{2n+1}}{\lambda_{2n+1}} H\dd\sigma,\ee 
where
\be \frac{\partial^2S_n}{\partial\sigma^2}-\lambda_n^2S_n-3\frac{\sigma\ped{dust}}{\sigma_0} H S_n=-2\lambda_n\sqrt{\frac{2\pi}{\tau_0}}\delta(\sigma).\ee 

As explained in the main text, we expect $M\ped{F}$ to reach asymptotically a finite value for $\tau_0\to\infty$. In this limit, the sum over $n$ can be converted to an integral over $\dd\lambda=\dd(\pi n/\tau_0)$ and we have
\begin{equation}
\label{eqn:MF-simplified}
M\ped{F}=\frac4\pi\int_0^\infty\dd\lambda\int_{-\infty}^{+\infty}S(\lambda,\sigma) H\dd\sigma,
\end{equation}
where
\begin{equation}
\label{eqn:S-simplified}
\frac{\partial^2S}{\partial\sigma^2}-(\lambda^2+3 H\epsilon)S=-\delta(\sigma),\qquad \epsilon\equiv\sigma\ped{dust}/\sigma_0.
\end{equation}
Given the absence of a particular feature at the \Lya\ frequency in the dust absorption profile, we assume $\epsilon$ to be independent of the frequency.

When $\epsilon=0$, we have $S=e^{-\lambda|\sigma|}/(2\lambda)$ and the contributions in the neighbourhood of $\lambda=0$ to \cref{eqn:MF-simplified} are logarithmically divergent. When $\epsilon\ne0$, however, the solution of \cref{eqn:S-simplified} deviates substantially from $e^{-\lambda|\sigma|}/(2\lambda)$ for sufficiently small $\lambda$, curing the divergence. Let us analyze the shape of the solution. At $\lambda=0$, we take the wing limit
\be  H\sim\frac{2^{1/3}a_v^{1/3}}{3\pi^{1/6}}\sigma^{-2/3}\ee 
and employing the WKB approximation\footnote{Given the equation $S''-k^2S=0$, the WKB approximation holds as long as $\frac1{k^2}\bigl|\frac{\dd k}{\dd\sigma}\bigr|\lesssim1$. In our case, $\frac1{\sqrt{\epsilon}|\sigma|^{2/3}}\frac{\pi^{1/12}}{3\cdot2^{1/6}a_v^{1/6}}\lesssim1$, meaning that the exponent in \cref{eqn:S-WKB} should be no more than $-1/2$. Therefore, the WKB approximation correctly captures the behaviour of the solution except in a neighbourhood of $\sigma=0$. In the wing approximation, the full solution for $\lambda=0$ is
\be
S=\frac{3\sigma^{1/2}}{2^{5/4}\Gamma(1/4)\beta^{3/4}}K_{\frac34}\bigl(\beta\sigma^{2/3}\bigr),\qquad \beta=\frac{3a_v^{1/6}\epsilon^{1/2}}{2^{5/6}\pi^{1/12}}
\ee
where $K_{\frac34}$ is the modified Bessel function of second kind.}:
\begin{equation}
\label{eqn:S-WKB}
S\sim\frac1{(3 H\epsilon)^{1/4}}e^{-\int_0^{|\sigma|}\sqrt{3 H\epsilon}\dd\sigma'}\approx\frac1{(3 H\epsilon)^{1/4}}e^{-\sqrt{\epsilon}\frac{3a_v^{1/6}}{2^{5/6}\pi^{1/12}}|\sigma|^{2/3}}.
\end{equation}
For sufficiently small $\lambda$, \cref{eqn:S-WKB} will still be a good approximation for the solution. The critical value of $\lambda$ for which significant deviations are expected is obtained comparing the typical ``decay length'' of \cref{eqn:S-WKB} with the one of $e^{-\lambda|\sigma|}/(2\lambda)$:
\begin{equation}
\label{eqn:lambda-c}
\lambda\ped{c}\biggl(\sqrt{\epsilon}\frac{3a_v^{1/6}}{2^{5/6}\pi^{1/12}}\biggr)^{-3/2}=1\implies\lambda\ped{c}=\epsilon^{3/4}\frac{3^{3/2}a_v^{1/4}}{2^{5/4}\pi^{1/8}}.
\end{equation}
For $\lambda<\lambda\ped{c}$, \cref{eqn:S-WKB} will be a good approximation. The contribution to $M\ped{F}$ is
\be M\ped{F}\ap{\lambda<\lambda\ped{c}}\sim\frac4\pi\int_0^{\lambda\ped{c}}\dd\lambda\int_{-\infty}^{+\infty}\frac1{3\epsilon}\biggl(\frac{\partial^2S}{\partial\sigma^2}+\delta(\sigma)\biggr)\dd\sigma\approx \frac{2^{3/4}3^{1/2}}{\pi^{9/8}}\frac{a_v^{1/4}}{\epsilon^{1/4}}.\ee 

In order to estimate the contribution for $\lambda>\lambda\ped{c}$, we can keep $\lambda^2$ instead of $3 H\epsilon$ in \cref{eqn:S-simplified} and get
\be M\ped{F}\ap{\lambda>\lambda\ped{c}}\sim\frac4\pi\sqrt{\frac23}\int_{\lambda\ped{c}}^\infty\dd\lambda\int_{-\infty}^{+\infty}\frac{e^{-\lambda|\sigma|}}{2\lambda}\dd x.\ee 
In the wing approximation, this is the same kind of integral done for the dust-free case and gives
\be M\ped{F}\ap{\lambda>\lambda\ped{c}}\sim\frac{2^{7/3}\cdot3}{\pi^{7/6}}\Gamma\biggl(\frac43\biggr)\frac{a_v^{1/3}}{\lambda\ped{c}^{1/3}}=\frac{2^{11/4}3^{1/2}}{\pi^{9/8}}\Gamma\biggl(\frac43\biggr)\frac{a_v^{1/4}}{\epsilon^{1/4}}.\ee 

Remarkably, we find the same dependence on $a_v$ and $\epsilon$ as obtained for $\lambda<\lambda\ped{c}$. We conclude that
\be \lim_{\tau_0\to\infty}M\ped{F}=\lim_{\tau_0\to\infty}\bigl(M\ped{F}\ap{\lambda<\lambda\ped{c}}+M\ped{F}\ap{\lambda>\lambda\ped{c}}\bigr)={\cal N}\times\frac{a_v^{1/4}}{\epsilon^{1/4}},\ee 
where we made the limit explicit for clarity and traded all numerical factors for a order-unity factor, ${\cal N}$.

\subsubsection{Constraining the numerical factor}

The numerical factors appearing in our computations above are an overestimate. We can see that this is the case by using the following lemma. Restricting to $\sigma>0$, consider
\be \begin{cases}
\frac{\partial^2S_1}{\partial\sigma^2}-f(\sigma)S_1=0\\
\frac{\partial^2S_2}{\partial\sigma^2}-g(\sigma)S_2=0\\
\frac{\partial S_1}{\partial\sigma}\Big|_0=\frac{\partial S_2}{\partial\sigma}\Big|_0=-1/2\\
\displaystyle\lim_{\sigma\to\infty}S_1=\lim_{\sigma\to\infty}S_2=0
\end{cases}\ee 
with $f(\sigma)>g(\sigma)$. Define $Y=S_1-S_2$. Then,
\be \frac{\partial^2Y}{\partial\sigma^2}=(f-g)S_1+gY>0,\qquad\frac{\partial Y}{\partial\sigma}\bigg|_0=0,\qquad \lim_{\sigma\to\infty}Y=0.
\ee 
If $Y(0)>0$, then $Y$ is forced to be a strictly non-decreasing function (since $Y'$ can only grow), violating the boundary condition at infinity. If $Y(0)<0$ and $Y>0$ at some point, there must exist a point where $Y=0$ and $Y'>0$ simultaneously. From that point on, $Y$ is again forced to be strictly non-decreasing, getting the same absurd conclusion. Therefore, $Y<0$ always.

In our estimate, we traded $\lambda^2+3 H\epsilon$ for either the first term (for $\lambda>\lambda\ped{c}$) or the second term (for $\lambda<\lambda\ped{c}$) only, so we can apply the lemma and conclude that $\cal N$ is less than the sum of the two numerical factors. Moreover, such conclusion is true for every choice of $\lambda\ped{c}$. Therefore, the best estimate is obtained rescaling $\lambda\ped{c}\to\alpha\lambda\ped{c}$ and minimizing with respect to $\alpha$. The minimum of
\be \frac{2^{3/4}3^{1/2}}{\pi^{9/8}}\biggl(\alpha+\frac{4\Gamma(4/3)}{\alpha^{1/3}}\biggr)\ee 
is obtained for
\be
\alpha=\biggl(\frac43\Gamma\biggl(\frac43\biggr)\biggr)^{3/4}
\ee
and we get
\be {\cal N}<\frac{2^{17/4}\Gamma(4/3)^{3/4}}{\pi^{9/8}3^{1/4}}\approx3.66.\ee

\begin{figure}
\includegraphics[width=0.48\textwidth]{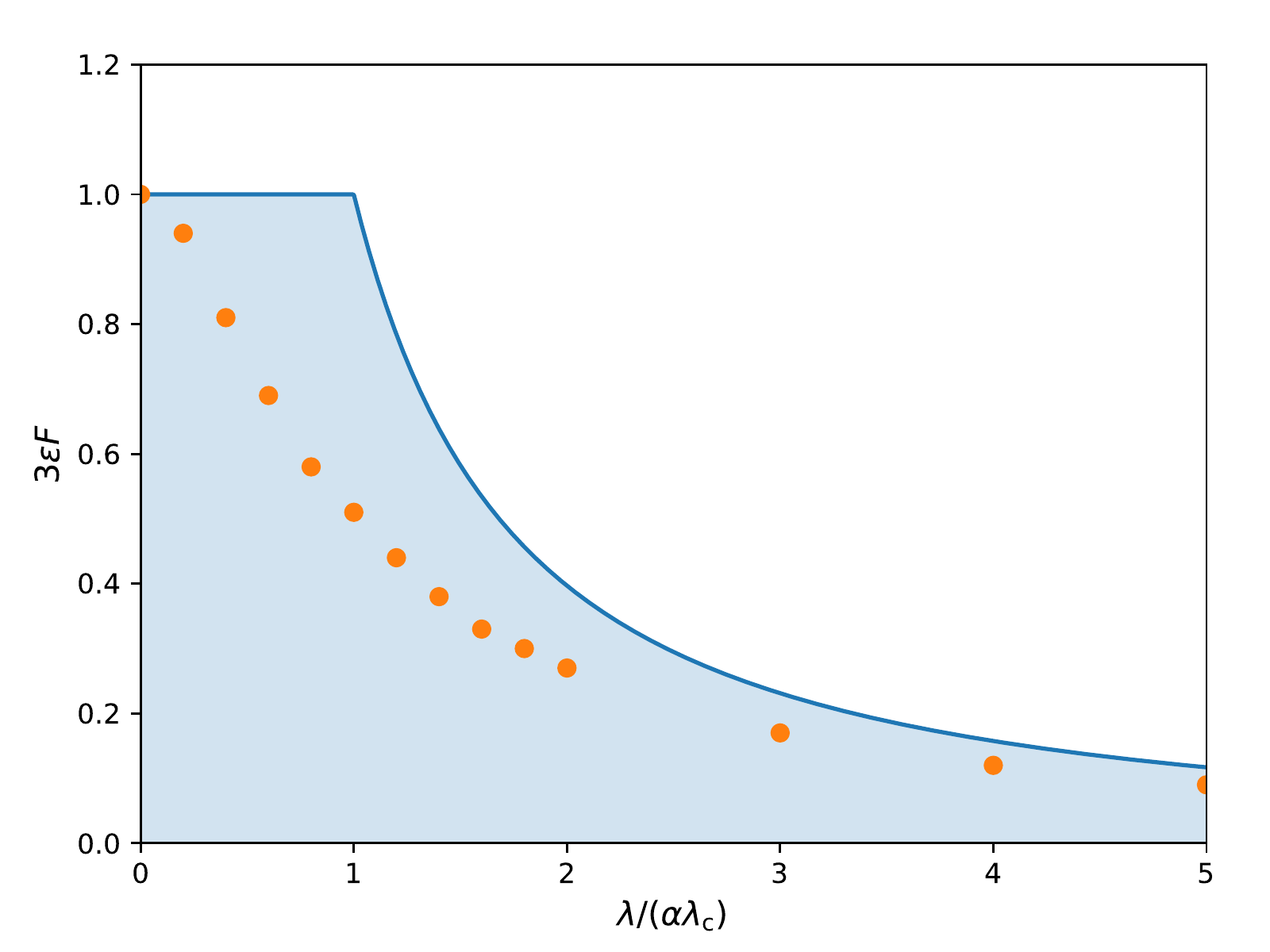}
\caption{Schematic representation of the integrals needed to compute $M\ped{F}$ (see \cref{eqn:def-F}). The blue surface (which extends indefinitely along the $x$ axis) has an area of 4 and corresponds to the overestimate presented in \ref{sec:dust-details}, which leads to ${\cal N}=3.66$. The orange dots are the results of several numerical integrations we performed in order to compute $F(\lambda)$, and thus $M\ped{F}$. The area beneath them is approximately $3.34$, leading to ${\cal N}=3.06$.
\label{fig:numerical-integration-dust} }
\end{figure}

To get a better estimate, we need to actually perform the integrals. Let us define
\begin{equation}
\label{eqn:def-F}
F(\lambda)=\int_{-\infty}^{+\infty}S(\lambda,\sigma) H\dd\sigma,\qquad\text{so that}\qquad M\ped{F}=\frac4\pi\int_0^\infty F\dd\lambda.
\end{equation}
Our previous overestimate was equivalent to
\be F(\lambda)\le\begin{cases}
\frac1{3\epsilon} & \text{for }\lambda<\alpha\lambda\ped{c}\\
\frac1{3\epsilon}\bigl(\frac{\alpha\lambda\ped{c}}\lambda\bigr)^{4/3} & \text{for }\lambda>\alpha\lambda\ped{c}
\end{cases}\ee 
which is depicted as the blue line in Fig.~\ref{fig:numerical-integration-dust}. Once the function $F(\lambda)$ is known, \cref{eqn:def-F} allows to determine $\cal N$ performing the integral over $\lambda$. In particular, with reference to the normalization of the axes in Fig.~\ref{fig:numerical-integration-dust},
\be
\mathcal N = \frac{2^{9/4}\Gamma(4/3)^{3/4}}{3^{1/4}\pi^{9/8}} \int_0^\infty3\epsilon F\dd\biggl(\frac\lambda{\alpha\lambda\ped{c}}\biggr).
\ee

We performed some numerical integrations to determine $F(\lambda)$. This is not easy, because of the vanishing boundary condition at infinity. The normalization of $S$ needs to be accurately chosen to get rid of the exponentially diverging solution. Our results are plotted as orange dots in Fig.~\ref{fig:numerical-integration-dust}. This allows a rough but easy trapezoidal estimate of the integral over $\dd\lambda$ to give $M\ped{F}$. Our final result is
\be \int_0^\infty3\epsilon F\dd\biggl(\frac\lambda{\alpha\lambda\ped{c}}\biggr)=3.34\implies {\cal N}=3.06,
\ee
with an uncertainty $< 10$\%.

\section*{Data availability}

The data underlying this article will be shared on reasonable request to the corresponding author.


\bibliographystyle{mnras}
\bibliography{Paper} 

\begin{thebibliography}{}
\makeatletter
\relax
\def\mn@urlcharsother{\let\do\@makeother \do\$\do\&\do\#\do\^\do\_\do\%\do\~}
\def\mn@doi{\begingroup\mn@urlcharsother \@ifnextchar [ {\mn@doi@}
  {\mn@doi@[]}}
\def\mn@doi@[#1]#2{\def\@tempa{#1}\ifx\@tempa\@empty \href
  {http://dx.doi.org/#2} {doi:#2}\else \href {http://dx.doi.org/#2} {#1}\fi
  \endgroup}
\def\mn@eprint#1#2{\mn@eprint@#1:#2::\@nil}
\def\mn@eprint@arXiv#1{\href {http://arxiv.org/abs/#1} {{\tt arXiv:#1}}}
\def\mn@eprint@dblp#1{\href {http://dblp.uni-trier.de/rec/bibtex/#1.xml}
  {dblp:#1}}
\def\mn@eprint@#1:#2:#3:#4\@nil{\def\@tempa {#1}\def\@tempb {#2}\def\@tempc
  {#3}\ifx \@tempc \@empty \let \@tempc \@tempb \let \@tempb \@tempa \fi \ifx
  \@tempb \@empty \def\@tempb {arXiv}\fi \@ifundefined
  {mn@eprint@\@tempb}{\@tempb:\@tempc}{\expandafter \expandafter \csname
  mn@eprint@\@tempb\endcsname \expandafter{\@tempc}}}

\bibitem[\protect\citeauthoryear{{Abe} \& {Yajima}}{{Abe} \&
  {Yajima}}{2018}]{Abe18}
{Abe} M.,  {Yajima} H.,  2018, \mn@doi [\mnras] {10.1093/mnrasl/sly018}, \href
  {https://ui.adsabs.harvard.edu/abs/2018MNRAS.475L.130A} {475, L130}

\bibitem[\protect\citeauthoryear{{Adams}}{{Adams}}{1972}]{Adams72}
{Adams} T.~F.,  1972, \mn@doi [\apj] {10.1086/151503}, \href
  {https://ui.adsabs.harvard.edu/abs/1972ApJ...174..439A} {174, 439}

\bibitem[\protect\citeauthoryear{{Adams}}{{Adams}}{1975}]{Adams:1975}
{Adams} T.~F.,  1975, \mn@doi [\apj] {10.1086/153891}, \href
  {https://ui.adsabs.harvard.edu/abs/1975ApJ...201..350A} {201, 350}

\bibitem[\protect\citeauthoryear{Ahn, Lee  \& Lee}{Ahn
  et~al.}{2002}]{Ahn:2001pz}
Ahn S.-H.,  Lee H.-W.,   Lee H.~M.,  2002, \mn@doi [Astrophys. J.]
  {10.1086/338497}, 567, 922

\bibitem[\protect\citeauthoryear{{Asplund}, {Grevesse}, {Sauval}  \&
  {Scott}}{{Asplund} et~al.}{2009}]{asplund2009}
{Asplund} M.,  {Grevesse} N.,  {Sauval} A.~J.,   {Scott} P.,  2009, \mn@doi
  [\araa] {10.1146/annurev.astro.46.060407.145222}, \href
  {http://adsabs.harvard.edu/abs/2009ARA%26A..47..481A} {47, 481}

\bibitem[\protect\citeauthoryear{{Ballesteros-Paredes}, {Hartmann},
  {V{\'a}zquez-Semadeni}, {Heitsch}  \&
  {Zamora-Avil{\'e}s}}{{Ballesteros-Paredes} et~al.}{2011}]{Ballesteros11}
{Ballesteros-Paredes} J.,  {Hartmann} L.~W.,  {V{\'a}zquez-Semadeni} E.,
  {Heitsch} F.,   {Zamora-Avil{\'e}s} M.~A.,  2011, \mn@doi [\mnras]
  {10.1111/j.1365-2966.2010.17657.x}, \href
  {https://ui.adsabs.harvard.edu/abs/2011MNRAS.411...65B} {411, 65}

\bibitem[\protect\citeauthoryear{Behrens, Dijkstra  \& Niemeyer}{Behrens
  et~al.}{2014}]{Behrens:2014bua}
Behrens C.,  Dijkstra M.,   Niemeyer J.,  2014, \mn@doi [Astron. Astrophys.]
  {10.1051/0004-6361/201322949}, 563, A77

\bibitem[\protect\citeauthoryear{{Behrens}, {Pallottini}, {Ferrara},
  {Gallerani}  \& {Vallini}}{{Behrens} et~al.}{2019}]{Behrens19}
{Behrens} C.,  {Pallottini} A.,  {Ferrara} A.,  {Gallerani} S.,   {Vallini} L.,
   2019, \mn@doi [\mnras] {10.1093/mnras/stz980}, \href
  {https://ui.adsabs.harvard.edu/abs/2019MNRAS.486.2197B} {486, 2197}

\bibitem[\protect\citeauthoryear{{Bithell}}{{Bithell}}{1990}]{Bithell90}
{Bithell} M.,  1990, \mnras, \href
  {https://ui.adsabs.harvard.edu/abs/1990MNRAS.244..738B} {244, 738}

\bibitem[\protect\citeauthoryear{{Bonilha}, {Ferch}, {Salpeter}, {Slater}  \&
  {Noerdlinger}}{{Bonilha} et~al.}{1979}]{1979ApJ...233..649B}
{Bonilha} J.~R.~M.,  {Ferch} R.,  {Salpeter} E.~E.,  {Slater} G.,
  {Noerdlinger} P.~D.,  1979, \mn@doi [\apj] {10.1086/157426}, \href
  {https://ui.adsabs.harvard.edu/abs/1979ApJ...233..649B} {233, 649}

\bibitem[\protect\citeauthoryear{{Ciardi} \& {Ferrara}}{{Ciardi} \&
  {Ferrara}}{2005}]{Ciardi05}
{Ciardi} B.,  {Ferrara} A.,  2005, \mn@doi [\ssr] {10.1007/s11214-005-3592-0},
  \href {https://ui.adsabs.harvard.edu/abs/2005SSRv..116..625C} {116, 625}

\bibitem[\protect\citeauthoryear{{Cox}}{{Cox}}{1985}]{Cox85}
{Cox} D.~P.,  1985, \mn@doi [\apj] {10.1086/162812}, \href
  {https://ui.adsabs.harvard.edu/abs/1985ApJ...288..465C} {288, 465}

\bibitem[\protect\citeauthoryear{{Dayal} \& {Ferrara}}{{Dayal} \&
  {Ferrara}}{2018}]{Dayal18}
{Dayal} P.,  {Ferrara} A.,  2018, \mn@doi [\physrep]
  {10.1016/j.physrep.2018.10.002}, \href
  {http://adsabs.harvard.edu/abs/2018PhR...780....1D} {780, 1}

\bibitem[\protect\citeauthoryear{{Dijkstra}}{{Dijkstra}}{2014}]{Dijkstra14}
{Dijkstra} M.,  2014, \mn@doi [\pasa] {10.1017/pasa.2014.33}, \href
  {https://ui.adsabs.harvard.edu/abs/2014PASA...31...40D} {31, e040}

\bibitem[\protect\citeauthoryear{{Dijkstra}}{{Dijkstra}}{2017}]{Dijkstra17}
{Dijkstra} M.,  2017, arXiv e-prints, \href
  {https://ui.adsabs.harvard.edu/abs/2017arXiv170403416D} {p. arXiv:1704.03416}

\bibitem[\protect\citeauthoryear{{Dijkstra} \& {Loeb}}{{Dijkstra} \&
  {Loeb}}{2008}]{Dijkstra08}
{Dijkstra} M.,  {Loeb} A.,  2008, \mn@doi [\mnras]
  {10.1111/j.1365-2966.2008.13920.x}, \href
  {https://ui.adsabs.harvard.edu/abs/2008MNRAS.391..457D} {391, 457}

\bibitem[\protect\citeauthoryear{{Dijkstra} \& {Loeb}}{{Dijkstra} \&
  {Loeb}}{2009}]{Dijkstra09}
{Dijkstra} M.,  {Loeb} A.,  2009, \mn@doi [\mnras]
  {10.1111/j.1365-2966.2009.14602.x}, \href
  {https://ui.adsabs.harvard.edu/abs/2009MNRAS.396..377D} {396, 377}

\bibitem[\protect\citeauthoryear{{Dijkstra}, {Haiman}  \& {Spaans}}{{Dijkstra}
  et~al.}{2006}]{Dijkstra06}
{Dijkstra} M.,  {Haiman} Z.,   {Spaans} M.,  2006, \mn@doi [\apj]
  {10.1086/506243}, \href
  {https://ui.adsabs.harvard.edu/abs/2006ApJ...649...14D} {649, 14}

\bibitem[\protect\citeauthoryear{Dijkstra, Prochaska, Ouchi  \& Hayes}{Dijkstra
  et~al.}{2019}]{bookDijkstra19}
Dijkstra M.,  Prochaska J.,  Ouchi M.,   Hayes M.,  2019, Lyman-alpha as an
  Astrophysical and Cosmological Tool: Saas-Fee Advanced Course 46. Swiss
  Society for Astrophysics and Astronomy, \mn@doi{10.1007/978-3-662-59623-4.
}

\bibitem[\protect\citeauthoryear{{Djorgovski} \& {Thompson}}{{Djorgovski} \&
  {Thompson}}{1992}]{Djorgovski92}
{Djorgovski} S.,  {Thompson} D.~J.,  1992, in {Barbuy} B.,  {Renzini} A.,  eds,
   IAU Symposium Vol. 149, The Stellar Populations of Galaxies. p.~337

\bibitem[\protect\citeauthoryear{Hansen \& Oh}{Hansen \&
  Oh}{2006}]{Hansen:2005pm}
Hansen M.,  Oh S.~P.,  2006, \mn@doi [Mon. Not. Roy. Astron. Soc.]
  {10.1111/j.1365-2966.2005.09870.x}, 367, 979

\bibitem[\protect\citeauthoryear{{Harrington}}{{Harrington}}{1973}]{Harrington73}
{Harrington} J.~P.,  1973, \mn@doi [\mnras] {10.1093/mnras/162.1.43}, \href
  {https://ui.adsabs.harvard.edu/abs/1973MNRAS.162...43H} {162, 43}

\bibitem[\protect\citeauthoryear{{Hayes}, {Runnholm}, {Gronke}  \&
  {Scarlata}}{{Hayes} et~al.}{2020}]{Hayes20}
{Hayes} M.~J.,  {Runnholm} A.,  {Gronke} M.,   {Scarlata} C.,  2020, arXiv
  e-prints, \href {https://ui.adsabs.harvard.edu/abs/2020arXiv200603232H} {p.
  arXiv:2006.03232}

\bibitem[\protect\citeauthoryear{{Herenz} et~al.,}{{Herenz}
  et~al.}{2019}]{MUSE19}
{Herenz} E.~C.,  et~al., 2019, \mn@doi [\aap] {10.1051/0004-6361/201834164},
  \href {https://ui.adsabs.harvard.edu/abs/2019A&A...621A.107H} {621, A107}

\bibitem[\protect\citeauthoryear{{Hill} \& {HETDEX Consortium}}{{Hill} \&
  {HETDEX Consortium}}{2016}]{Hetdex16}
{Hill} G.~J.,  {HETDEX Consortium} 2016, in {Skillen} I.,  {Balcells} M.,
  {Trager} S.,  eds,  Astronomical Society of the Pacific Conference Series
  Vol. 507, Multi-Object Spectroscopy in the Next Decade: Big Questions, Large
  Surveys, and Wide Fields. p.~393

\bibitem[\protect\citeauthoryear{{Hopkins}, {Grudi{\'c}}, {Wetzel},
  {Kere{\v{s}}}, {Faucher-Gigu{\`e}re}, {Ma}, {Murray}  \& {Butcher}}{{Hopkins}
  et~al.}{2020}]{Hopkins20}
{Hopkins} P.~F.,  {Grudi{\'c}} M.~Y.,  {Wetzel} A.,  {Kere{\v{s}}} D.,
  {Faucher-Gigu{\`e}re} C.-A.,  {Ma} X.,  {Murray} N.,   {Butcher} N.,  2020,
  \mn@doi [\mnras] {10.1093/mnras/stz3129}, \href
  {https://ui.adsabs.harvard.edu/abs/2020MNRAS.491.3702H} {491, 3702}

\bibitem[\protect\citeauthoryear{{Hu}, {Cowie}, {Barger}, {Capak}, {Kakazu}  \&
  {Trouille}}{{Hu} et~al.}{2010}]{Hu10}
{Hu} E.~M.,  {Cowie} L.~L.,  {Barger} A.~J.,  {Capak} P.,  {Kakazu} Y.,
  {Trouille} L.,  2010, \mn@doi [\apj] {10.1088/0004-637X/725/1/394}, \href
  {https://ui.adsabs.harvard.edu/abs/2010ApJ...725..394H} {725, 394}

\bibitem[\protect\citeauthoryear{{Hummer}}{{Hummer}}{1962}]{Hummer:1962}
{Hummer} D.~G.,  1962, \mn@doi [\mnras] {10.1093/mnras/125.1.21}, \href
  {https://ui.adsabs.harvard.edu/abs/1962MNRAS.125...21H} {125, 21}

\bibitem[\protect\citeauthoryear{{Jung} et~al.,}{{Jung} et~al.}{2020}]{Jung20}
{Jung} I.,  et~al., 2020, arXiv e-prints, \href
  {https://ui.adsabs.harvard.edu/abs/2020arXiv200910092J} {p. arXiv:2009.10092}

\bibitem[\protect\citeauthoryear{{Kashikawa} et~al.,}{{Kashikawa}
  et~al.}{2011}]{Kashikawa11}
{Kashikawa} N.,  et~al., 2011, \mn@doi [\apj] {10.1088/0004-637X/734/2/119},
  \href {https://ui.adsabs.harvard.edu/abs/2011ApJ...734..119K} {734, 119}

\bibitem[\protect\citeauthoryear{{Kimm}, {Haehnelt}, {Blaizot}, {Katz},
  {Michel-Dansac}, {Garel}, {Rosdahl}  \& {Teyssier}}{{Kimm}
  et~al.}{2018}]{Kimm18}
{Kimm} T.,  {Haehnelt} M.,  {Blaizot} J.,  {Katz} H.,  {Michel-Dansac} L.,
  {Garel} T.,  {Rosdahl} J.,   {Teyssier} R.,  2018, \mn@doi [\mnras]
  {10.1093/mnras/sty126}, \href
  {https://ui.adsabs.harvard.edu/abs/2018MNRAS.475.4617K} {475, 4617}

\bibitem[\protect\citeauthoryear{{Kimm}, {Blaizot}, {Garel}, {Michel-Dansac},
  {Katz}, {Rosdahl}, {Verhamme}  \& {Haehnelt}}{{Kimm} et~al.}{2019}]{Kimm19}
{Kimm} T.,  {Blaizot} J.,  {Garel} T.,  {Michel-Dansac} L.,  {Katz} H.,
  {Rosdahl} J.,  {Verhamme} A.,   {Haehnelt} M.,  2019, \mn@doi [\mnras]
  {10.1093/mnras/stz989}, \href
  {https://ui.adsabs.harvard.edu/abs/2019MNRAS.486.2215K} {486, 2215}

\bibitem[\protect\citeauthoryear{{Krumholz}, {Dekel}  \& {McKee}}{{Krumholz}
  et~al.}{2012}]{krumholz2012}
{Krumholz} M.~R.,  {Dekel} A.,   {McKee} C.~F.,  2012, \mn@doi [\apj]
  {10.1088/0004-637X/745/1/69}, \href
  {http://adsabs.harvard.edu/abs/2012ApJ...745...69K} {745, 69}

\bibitem[\protect\citeauthoryear{Lao \& Smith}{Lao \&
  Smith}{2020}]{Lao:2020ptq}
Lao B.-X.,  Smith A.,  2020, \mn@doi [Mon. Not. Roy. Astron. Soc.]
  {10.1093/mnras/staa2198}, 497, 3925

\bibitem[\protect\citeauthoryear{{Laursen}, {Sommer-Larsen}, {Milvang-Jensen},
  {Fynbo}  \& {Razoumov}}{{Laursen} et~al.}{2019}]{Laursen19}
{Laursen} P.,  {Sommer-Larsen} J.,  {Milvang-Jensen} B.,  {Fynbo} J. P.~U.,
  {Razoumov} A.~O.,  2019, \mn@doi [\aap] {10.1051/0004-6361/201833645}, \href
  {https://ui.adsabs.harvard.edu/abs/2019A&A...627A..84L} {627, A84}

\bibitem[\protect\citeauthoryear{{Li}, {Gu}, {Yajima}, {Zhu}  \& {Maji}}{{Li}
  et~al.}{2020}]{Li20}
{Li} Y.,  {Gu} M.~F.,  {Yajima} H.,  {Zhu} Q.,   {Maji} M.,  2020, \mn@doi
  [\mnras] {10.1093/mnras/staa733}, \href
  {https://ui.adsabs.harvard.edu/abs/2020MNRAS.494.1919L} {494, 1919}

\bibitem[\protect\citeauthoryear{McKee \& Tan}{McKee \&
  Tan}{2008}]{McKee:2007yx}
McKee C.~F.,  Tan J.~C.,  2008, \mn@doi [Astrophys. J.] {10.1086/587434}, 681,
  771

\bibitem[\protect\citeauthoryear{{Mesinger}, {Aykutalp}, {Vanzella},
  {Pentericci}, {Ferrara}  \& {Dijkstra}}{{Mesinger} et~al.}{2015}]{Mesinger15}
{Mesinger} A.,  {Aykutalp} A.,  {Vanzella} E.,  {Pentericci} L.,  {Ferrara} A.,
    {Dijkstra} M.,  2015, \mn@doi [\mnras] {10.1093/mnras/stu2089}, \href
  {https://ui.adsabs.harvard.edu/abs/2015MNRAS.446..566M} {446, 566}

\bibitem[\protect\citeauthoryear{{Michel-Dansac}, {Blaizot}, {Garel},
  {Verhamme}, {Kimm}  \& {Trebitsch}}{{Michel-Dansac} et~al.}{2020}]{Michel20}
{Michel-Dansac} L.,  {Blaizot} J.,  {Garel} T.,  {Verhamme} A.,  {Kimm} T.,
  {Trebitsch} M.,  2020, \mn@doi [\aap] {10.1051/0004-6361/201834961}, \href
  {https://ui.adsabs.harvard.edu/abs/2020A&A...635A.154M} {635, A154}

\bibitem[\protect\citeauthoryear{{Neufeld}}{{Neufeld}}{1990}]{Neufeld90}
{Neufeld} D.~A.,  1990, \mn@doi [\apj] {10.1086/168375}, \href
  {https://ui.adsabs.harvard.edu/abs/1990ApJ...350..216N} {350, 216}

\bibitem[\protect\citeauthoryear{Oh \& Haiman}{Oh \& Haiman}{2002}]{Oh:2001ex}
Oh S.~P.,  Haiman Z.,  2002, \mn@doi [Astrophys. J.] {10.1086/339393}, 569, 558

\bibitem[\protect\citeauthoryear{{Osterbrock}}{{Osterbrock}}{1962}]{Osterbrock:1962}
{Osterbrock} D.~E.,  1962, \mn@doi [\apj] {10.1086/147258}, \href
  {https://ui.adsabs.harvard.edu/abs/1962ApJ...135..195O} {135, 195}

\bibitem[\protect\citeauthoryear{{Osterbrock} \& {Ferland}}{{Osterbrock} \&
  {Ferland}}{2006}]{Osterbrock06}
{Osterbrock} D.~E.,  {Ferland} G.~J.,  2006, {Astrophysics of gaseous nebulae
  and active galactic nuclei}

\bibitem[\protect\citeauthoryear{{Ouchi} et~al.,}{{Ouchi}
  et~al.}{2009}]{Ouchi09}
{Ouchi} M.,  et~al., 2009, \mn@doi [\apj] {10.1088/0004-637X/696/2/1164}, \href
  {https://ui.adsabs.harvard.edu/abs/2009ApJ...696.1164O} {696, 1164}

\bibitem[\protect\citeauthoryear{{Ouchi} et~al.,}{{Ouchi}
  et~al.}{2018}]{Ouchi18}
{Ouchi} M.,  et~al., 2018, \mn@doi [\pasj] {10.1093/pasj/psx074}, \href
  {https://ui.adsabs.harvard.edu/abs/2018PASJ...70S..13O} {70, S13}

\bibitem[\protect\citeauthoryear{Ouchi, Ono  \& Shibuya}{Ouchi
  et~al.}{2020}]{Ouchi:2020zce}
Ouchi M.,  Ono Y.,   Shibuya T.,  2020, \mn@doi [Ann. Rev. Astron. Astrophys.]
  {10.1146/annurev-astro-032620-021859}, 58, 617

\bibitem[\protect\citeauthoryear{{Pallottini}, {Ferrara}, {Gallerani},
  {Vallini}, {Maiolino}  \& {Salvadori}}{{Pallottini}
  et~al.}{2017}]{pallottini2017}
{Pallottini} A.,  {Ferrara} A.,  {Gallerani} S.,  {Vallini} L.,  {Maiolino} R.,
    {Salvadori} S.,  2017, \mn@doi [\mnras] {10.1093/mnras/stw2847}, \href
  {http://adsabs.harvard.edu/abs/2017MNRAS.465.2540P} {465, 2540}

\bibitem[\protect\citeauthoryear{{Partridge} \& {Peebles}}{{Partridge} \&
  {Peebles}}{1967}]{Partridge67}
{Partridge} R.~B.,  {Peebles} P.~J.~E.,  1967, \mn@doi [\apj] {10.1086/149079},
  \href {https://ui.adsabs.harvard.edu/abs/1967ApJ...147..868P} {147, 868}

\bibitem[\protect\citeauthoryear{{Pentericci} et~al.,}{{Pentericci}
  et~al.}{2011}]{Pentericci11}
{Pentericci} L.,  et~al., 2011, \mn@doi [\apj] {10.1088/0004-637X/743/2/132},
  \href {https://ui.adsabs.harvard.edu/abs/2011ApJ...743..132P} {743, 132}

\bibitem[\protect\citeauthoryear{{R{\'e}my-Ruyer} et~al.,}{{R{\'e}my-Ruyer}
  et~al.}{2014}]{Remy14}
{R{\'e}my-Ruyer} A.,  et~al., 2014, \mn@doi [\aap]
  {10.1051/0004-6361/201322803}, \href
  {https://ui.adsabs.harvard.edu/abs/2014A&A...563A..31R} {563, A31}

\bibitem[\protect\citeauthoryear{{Rhoads}, {Malhotra}, {Dey}, {Stern},
  {Spinrad}  \& {Jannuzi}}{{Rhoads} et~al.}{2000}]{Rhoads00}
{Rhoads} J.~E.,  {Malhotra} S.,  {Dey} A.,  {Stern} D.,  {Spinrad} H.,
  {Jannuzi} B.~T.,  2000, \mn@doi [\apjl] {10.1086/317874}, \href
  {https://ui.adsabs.harvard.edu/abs/2000ApJ...545L..85R} {545, L85}

\bibitem[\protect\citeauthoryear{{Schaerer}}{{Schaerer}}{2003}]{Schaerer03}
{Schaerer} D.,  2003, \mn@doi [\aap] {10.1051/0004-6361:20021525}, \href
  {https://ui.adsabs.harvard.edu/abs/2003A&A...397..527S} {397, 527}

\bibitem[\protect\citeauthoryear{{Seon} \& {Kim}}{{Seon} \&
  {Kim}}{2020}]{Seon:2020}
{Seon} K.-i.,  {Kim} C.-G.,  2020, \mn@doi [\apjs] {10.3847/1538-4365/aba2d6},
  \href {https://ui.adsabs.harvard.edu/abs/2020ApJS..250....9S} {250, 9}

\bibitem[\protect\citeauthoryear{{Shibuya} et~al.,}{{Shibuya}
  et~al.}{2018}]{Shibuya18}
{Shibuya} T.,  et~al., 2018, \mn@doi [\pasj] {10.1093/pasj/psx122}, \href
  {https://ui.adsabs.harvard.edu/abs/2018PASJ...70S..14S} {70, S14}

\bibitem[\protect\citeauthoryear{{Shibuya}, {Ouchi}, {Harikane}  \&
  {Nakajima}}{{Shibuya} et~al.}{2019}]{Shibuya2019}
{Shibuya} T.,  {Ouchi} M.,  {Harikane} Y.,   {Nakajima} K.,  2019, \mn@doi
  [\apj] {10.3847/1538-4357/aaf64b}, \href
  {http://adsabs.harvard.edu/abs/2019ApJ...871..164S} {871, 164}

\bibitem[\protect\citeauthoryear{{Smith}, {Bromm}  \& {Loeb}}{{Smith}
  et~al.}{2016}]{Smith:2006}
{Smith} A.,  {Bromm} V.,   {Loeb} A.,  2016, \mn@doi [\mnras]
  {10.1093/mnras/stw1129}, \href
  {https://ui.adsabs.harvard.edu/abs/2016MNRAS.460.3143S} {460, 3143}

\bibitem[\protect\citeauthoryear{{Smith}, {Bromm}  \& {Loeb}}{{Smith}
  et~al.}{2017a}]{Smith17}
{Smith} A.,  {Bromm} V.,   {Loeb} A.,  2017a, \mn@doi [\mnras]
  {10.1093/mnras/stw2591}, \href
  {https://ui.adsabs.harvard.edu/abs/2017MNRAS.464.2963S} {464, 2963}

\bibitem[\protect\citeauthoryear{Smith, Becerra, Bromm  \& Hernquist}{Smith
  et~al.}{2017b}]{Smith:2017xwu}
Smith A.,  Becerra F.,  Bromm V.,   Hernquist L.,  2017b, \mn@doi [Mon. Not.
  Roy. Astron. Soc.] {10.1093/mnras/stx1993}, 472, 205

\bibitem[\protect\citeauthoryear{{Smith}, {Tsang}, {Bromm}  \&
  {Milosavljevi{\'c}}}{{Smith} et~al.}{2018}]{Smith18}
{Smith} A.,  {Tsang} B. T.~H.,  {Bromm} V.,   {Milosavljevi{\'c}} M.,  2018,
  \mn@doi [\mnras] {10.1093/mnras/sty1509}, \href
  {https://ui.adsabs.harvard.edu/abs/2018MNRAS.479.2065S} {479, 2065}

\bibitem[\protect\citeauthoryear{{Smith}, {Ma}, {Bromm}, {Finkelstein},
  {Hopkins}, {Faucher-Gigu{\`e}re}  \& {Kere{\v{s}}}}{{Smith}
  et~al.}{2019}]{Smith19}
{Smith} A.,  {Ma} X.,  {Bromm} V.,  {Finkelstein} S.~L.,  {Hopkins} P.~F.,
  {Faucher-Gigu{\`e}re} C.-A.,   {Kere{\v{s}}} D.,  2019, \mn@doi [\mnras]
  {10.1093/mnras/sty3483}, \href
  {https://ui.adsabs.harvard.edu/abs/2019MNRAS.484...39S} {484, 39}

\bibitem[\protect\citeauthoryear{{Sobacchi} \& {Mesinger}}{{Sobacchi} \&
  {Mesinger}}{2015}]{Sobacchi15}
{Sobacchi} E.,  {Mesinger} A.,  2015, \mn@doi [\mnras] {10.1093/mnras/stv1751},
  \href {https://ui.adsabs.harvard.edu/abs/2015MNRAS.453.1843S} {453, 1843}

\bibitem[\protect\citeauthoryear{{Sommovigo}, {Ferrara}, {Pallottini},
  {Carniani}, {Gallerani}  \& {Decataldo}}{{Sommovigo}
  et~al.}{2020}]{Sommovigo20}
{Sommovigo} L.,  {Ferrara} A.,  {Pallottini} A.,  {Carniani} S.,  {Gallerani}
  S.,   {Decataldo} D.,  2020, \mn@doi [\mnras] {10.1093/mnras/staa1959}, \href
  {https://ui.adsabs.harvard.edu/abs/2020MNRAS.497..956S} {497, 956}

\bibitem[\protect\citeauthoryear{{Tan} \& {McKee}}{{Tan} \&
  {McKee}}{2003}]{Tan03}
{Tan} J.~C.,  {McKee} C.~F.,  2003, in {Holt} S.~H.,  {Reynolds} C.~S.,  eds,
  American Institute of Physics Conference Series Vol. 666, The Emergence of
  Cosmic Structure. pp 93--96 (\mn@eprint {arXiv} {astro-ph/0212283}),
  \mn@doi{10.1063/1.1581776}

\bibitem[\protect\citeauthoryear{{Taniguchi} et~al.,}{{Taniguchi}
  et~al.}{2005}]{Taniguchi05}
{Taniguchi} Y.,  et~al., 2005, \mn@doi [\pasj] {10.1093/pasj/57.1.165}, \href
  {https://ui.adsabs.harvard.edu/abs/2005PASJ...57..165T} {57, 165}

\bibitem[\protect\citeauthoryear{Unno}{Unno}{1952}]{unno1952note}
Unno W.,  1952, Publications of the Astronomical Society of Japan, 4, 100

\bibitem[\protect\citeauthoryear{{Weinberger}, {Haehnelt}  \&
  {Kulkarni}}{{Weinberger} et~al.}{2019}]{Weinberger19}
{Weinberger} L.~H.,  {Haehnelt} M.~G.,   {Kulkarni} G.,  2019, \mn@doi [\mnras]
  {10.1093/mnras/stz481}, \href
  {https://ui.adsabs.harvard.edu/abs/2019MNRAS.485.1350W} {485, 1350}

\bibitem[\protect\citeauthoryear{{Weingartner} \& {Draine}}{{Weingartner} \&
  {Draine}}{2001}]{weingartner2001}
{Weingartner} J.~C.,  {Draine} B.~T.,  2001, \mn@doi [\apj] {10.1086/318651},
  \href {http://adsabs.harvard.edu/abs/2001ApJ...548..296W} {548, 296}

\bibitem[\protect\citeauthoryear{{Whitler}, {Mason}, {Ren}, {Dijkstra},
  {Mesinger}, {Pentericci}, {Trenti}  \& {Treu}}{{Whitler}
  et~al.}{2020}]{Whitler20}
{Whitler} L.~R.,  {Mason} C.~A.,  {Ren} K.,  {Dijkstra} M.,  {Mesinger} A.,
  {Pentericci} L.,  {Trenti} M.,   {Treu} T.,  2020, \mn@doi [\mnras]
  {10.1093/mnras/staa1178}, \href
  {https://ui.adsabs.harvard.edu/abs/2020MNRAS.495.3602W} {495, 3602}

\makeatother
\end{thebibliography}


\bsp	
\label{lastpage}
\end{document}